\documentclass[prd,twocolumn,showpacs,nofootinbib]{revtex4}
\usepackage{times}
\usepackage{natbib}
\usepackage{epsfig}

\newcommand{\OM}{\Omega_m}
\newcommand{\OB}{\Omega_b}

\newcommand{\rms}{\sigma_8}
\newcommand{\up}[1]{{\rm #1}}

\newcommand{\mpc}{{\rm Mpc}}

\newcommand{\hmpc}{{h^{-1}\mpc}}

\newcommand{\hmpci}{{h\mpc^{-1}}}
\newcommand{\mpci}{{\mpc^{-1}}}

\newcommand{\ang}{\hat\nabla}

\newcommand{\bdv}[1]{{\bf #1}}
\newcommand{\beeq}{\vspace{10pt}\begin{equation}} 
\newcommand{\eneq}{\vspace{10pt}\end{equation}}
\newcommand{\bear}{\vspace{10pt}\begin{eqnarray}}
\newcommand{\enar}{\end{eqnarray}{\\\\ \noindent}}

\newcommand{\Vang}{\bdv{\hat n}}
\newcommand{\Sang}{\bdv{\hat s}}
\newcommand{\zobs}{z_\up{obs}}
\newcommand{\ztobs}{\tilde z}
\newcommand{\tobs}{\tilde\theta}
\newcommand{\pobs}{\tilde\phi}
\newcommand{\dobs}{\delta_\up{obs}}
\newcommand{\nobs}{n_\up{obs}}
\newcommand{\ntobs}{\tilde n}
\newcommand{\fobs}{f_\up{obs}}

\newcommand{\dMB}{\delta_\up{mb}}
\newcommand{\devo}{\delta_\up{evo}}
\newcommand{\zdist}{\delta_\up{z}}
\newcommand{\rtd}{r_3}

\newcommand{\xim}{\xi_m}
\newcommand{\Pm}{P_m}
\newcommand{\xiMB}{\xi_\up{mb}}
\newcommand{\xiobs}{\xi_\up{obs}}
\newcommand{\xievo}{\xi_\up{evo}}
\newcommand{\xint}{\xi_\up{int}}
\newcommand{\xidMB}{\xi_\up{\delta~\up{mb}}}
\newcommand{\xizz}{\xi_\up{zz}}
\newcommand{\xidz}{\xi_\up{\delta\up{z}}}
\newcommand{\xizd}{\xi_\up{\up{z}\delta}}
\newcommand{\clcor}[1]{C_l^{#1}}

\begin{document}

\title{Complete treatment of galaxy two-point statistics: \\
gravitational lensing effects and redshift-space distortions}

\author{Jaiyul Yoo}
\altaffiliation{Electronic address: jyoo@cfa.harvard.edu}
\affiliation{Harvard-Smithsonian Center for Astrophysics, Harvard University,
60 Garden Street, Cambridge, MA 02138}

\date{\today}

\begin{abstract}
We present a coherent theoretical framework for computing gravitational
lensing effects and redshift-space distortions in an inhomogeneous universe
and investigate their impacts on galaxy two-point statistics.
Adopting the linearized Friedmann-Lema\^\i tre-Robertson-Walker metric,
we derive the gravitational lensing and the 
generalized Sachs-Wolfe effects that
include the weak lensing distortion, magnification, and time delay effects,
and the redshift-space distortion, Sachs-Wolfe, and integrated Sachs-Wolfe 
effects, respectively. 
Based on this framework, we first compute their effects on observed source 
fluctuations, separating them 
as two physically distinct origins: the volume effect that involves
the change of volume and is always present in galaxy two-point statistics,
and the source effect that depends on the intrinsic properties of source 
populations. Then
we identify several terms that are ignored in the
standard method, and we compute the observed 
galaxy two-point statistics, an ensemble average of all the combinations
of the intrinsic source fluctuations and the additional contributions from
the gravitational lensing and the generalized Sachs-Wolfe effects.
This unified treatment of galaxy two-point statistics clarifies the relation
of the gravitational lensing and the generalized Sachs-Wolfe effects
to the metric perturbations and the underlying matter fluctuations.
For near future dark energy surveys, we compute 
additional contributions to the observed galaxy two-point statistics
and analyze their impact on the anisotropic structure.
Thorough theoretical modeling of galaxy two-point statistics would be
not only necessary to analyze precision measurements from upcoming dark energy 
surveys, but also provide further 
discriminatory power in understanding the underlying physical mechanisms.
\end{abstract}


\maketitle

\section{Introduction}
\label{sec:intro}
The standard inflationary models with a single inflaton potential predict
a nearly perfect Gaussian spectrum of primordial fluctuations 
\citep{BASTTU83,STARO82,HAWKI82,GUPI82,MUFEBR92}. Two-point
statistics, correlation function in real space and power spectrum in Fourier
space, constitutes a complete description of Gaussian random fields, and it
has been widely used to understand the physics of the early universe from
measurements of the cosmic microwave background and large-scale structure.
The recent discovery \citep{RIFIET98,PEADET99}
of the late time acceleration of the universe has spurred
extensive investigations of a mysterious
energy component with negative pressure, dubbed  dark energy. 
Observationally, 
upcoming dark energy surveys will measure galaxy two-point statistics with
unprecedented precision from millions of galaxies, constraining the expansion
history and the spatial curvature of the universe. Consequently, accurate 
theoretical modeling of galaxy two-point statistics would be 
crucial to take full advantage of the promise that these
future surveys will deliver.

In achieving this goal, complications arise notably from the nonlinear
evolution of matter and scale-dependence of galaxy bias. 
In this paper we limit ourselves to the linear bias model \citep{KAISE84} 
and study the linear theory predictions and its corrections, considering that
recent attention has been paid to measuring galaxy two-point statistics
in the linear regime (e.g., \citep{EIBLET05,TEEIST06,PENIET07,PASCET07}).
However, measurement precision is often highest on nonlinear scales, and
proper modeling of galaxy bias on nonlinear scales can substantially increase 
the leverage to constrain the underlying physics (see, e.g.,
\citep{JIMOBO98,SELJA00,MAFR00,PESM00,SCSHET01,BEWE02,COSH02}).

Further complication arises from the distortion of
redshift-space structure by peculiar velocities, which results in anisotropy
from otherwise isotropic two-point statistics \citep{KAISE87,HAMIL92}.
The standard practice is to analyze the angle-averaged correlation function 
or power spectrum, or to construct a linear combination of their multipole 
components, suppressing the angular dependence of two-point statistics. 
However, analyzing the full anisotropic structure, though observationally 
challenging, can utilize additional information that is lost to some degree
in the standard practice \citep{MATSU04,SEEI07,OKMAET08,GACAHU08}.

Gravitational lensing, often assumed to be negligible in galaxy two-point
statistics, deflects the 
propagation of light rays, displacing the position of observed
galaxies, and it alters the unit area on the sky and magnifies
the observed flux, changing the observed number density of galaxies.
The former effect on two-point statistics is to convolve it with the power 
spectrum of the lensing potential, smoothing out the features in galaxy
two-point statistics \citep{SELJA96}.
The latter effect, known as the magnification bias \citep{NARAY89},
is often used to measure the galaxy-matter cross-correlation function
from two source populations separated by large line-of-sight distance
\citep{SCMEET05,BLPOET06}. Recent work \citep{MATSU00,VADOET07,HUGALO07}
showed that these effects on galaxy two-point statistics are
non-negligible at the level of accuracy adequate for upcoming dark energy
surveys.

However, it is unclear whether this list of additional contributions
on galaxy two-point statistics is exhaustive, and what are the contribution
terms that are ignored
in the standard method but need to be considered if higher accuracy is 
dictated by observations. 
Here we present a coherent theoretical framework for computing 
gravitational lensing effects and redshift-space distortions, and investigate
their impacts on galaxy two-point statistics in an inhomogeneous universe.
Our treatment generalizes the early work \citep{MATSU00}
and complements the recent work \citep{VADOET07,HUGALO07},
providing a unified description of galaxy two-point statistics.
However, we emphasize that these effects naturally arise from metric 
perturbations in our approach, comprising a complete and exhaustive set of
additional (linear order) contributions to galaxy two-point statistic.

The rest of this paper is organized as follows.
In Sec.~\ref{sec:formalism}, we describe our notation for the 
Friedmann-Lema{\^\i}tre-Robertson-Walker (FLRW) metric
and derive the gravitational lensing and the generalized Sachs-Wolfe effects.
In Sec.~\ref{sec:density}, we study their impacts on source galaxy 
fluctuations and discuss their correspondence to the standard redshift-space
distortion and gravitational lensing effect. In Sec.~\ref{ssec:two},
we derive the observed galaxy two-point statistics
in real space and in Fourier space, and we compare the effects
of each contribution term on the observed galaxy two-point statistics 
in Sec.~\ref{ssec:com}. We conclude in Sec.~\ref{sec:discuss} with a
discussion of the further improvement of our approach.

\section{Formalism}
\label{sec:formalism}
Here we describe our notation for a background metric in an inhomogeneous
universe and derive
governing equations for non-relativistic matter in Sec.~\ref{ssec:metric}.
Combining these with photon geodesic equations, we derive the generalized
Sachs-Wolfe (Sec.~\ref{ssec:sachs}) and the gravitational lensing 
(Sec.~\ref{ssec:lensing}) effects, developing a coherent
framework for describing how matter fluctuations affect observable quantities.

\subsection{Metric and Perturbations}
\label{ssec:metric}
We assume that the homogeneous and isotropic background of the universe is
described by the Friedmann-Lema\^\i tre-Robertson-Walker (FLRW)
metric and its inhomogeneous part is represented by the perturbations for the
cosmological fluids and the spacetime geometry:
\beeq
ds^2=-a^2(\eta)\left[1+2\psi\right]d\eta^2+a^2(\eta)\left[1+2\phi\right]
g_{\alpha\beta}^{(3)}dx^\alpha dx^\beta,
\label{eq:metric}
\eneq
with the metric tensor for a three-space of constant spatial curvature
$K=-H^2_0(1-\Omega_0)$,
\beeq
g_{\alpha\beta}^{(3)}dx^\alpha dx^\beta=d\chi^2+r^2(\chi)d\Omega^2,
\eneq
where $a(\eta)$ is the scale factor for the expansion of the background
as a function of the conformal time $\eta$, and the comoving angular diameter
distance is
$r(\chi)=K^{-1/2}\sin(\sqrt{K}\chi)$ for a closed universe $K>0$ and
$(-K)^{-1/2}\sinh(\sqrt{-K}\chi)$ for an open universe
$K<0$, where $\chi$ is the comoving line-of-sight distance.
The flat limit can be obtained as $K\rightarrow0$.
We will denote the covariant derivative of a three-tensor
with respect to $g^{(3)}_{\alpha\beta}$ as a vertical bar and the covariant
derivative in the spacetime metric as a semicolon in the following.
Here Latin indices represent 4D space-time components, and
Greek indices run from~1 to~3, representing the spatial part of the metric. 
Throughout the paper, we set the speed of light $c\equiv1$ 

We express the perturbations in the conformal Newtonian gauge, where $\psi$
and $\phi$ correspond to the intuitive physical quantities, i.e., Newtonian
potential and Newtonian
curvature. This choice of gauge condition leaves no residual
degree of freedom up to the first-order in
perturbations. Here we only consider scalar perturbations, as primordial
vector perturbations decay quickly in a universe with ordinary components
and the current upper limit on tensor perturbations is order of magnitude
smaller than the amplitude of scalar perturbations 
(e.g., \citep{TESTET06,SPBEET07,KODUET08})

Given the stress energy tensor $T^{ab}$ of cosmological components,
the evolution of the matter and metric perturbations is governed by the
Einstein equations $G_{ab}=8\pi GT_{ab}$, and the Bianchi identities
$T^{ab}{_{;b}}=0$ guarantee the conservation of energy and momentum
(e.g., \citep{BARDE80,KOSA84,MUFEBR92,HWNO02,HU04}). 
Current cosmological observations favor a universe dominated by   
dark energy, but with non-relativistic matter as the major source of metric
perturbations. In this universe, the scalar Einstein equations are
\bear
(k^2-3K)~\phi&=&{3H_0^2\over2}~\OM~\left[{\delta\over a}+3H~{v\over k}\right],
\\
\psi&=&-~\phi,
\enar
where $\delta$ is the density perturbation in non-relativistic matter.
The Hubble parameter is $H=\dot a/a$, where the overdot
is the derivative with respect to time, $dt=a~d\eta$. The matter density and
the Hubble parameters at the present day $a_0$ are denoted as $\OM$ and
$H_0$, respectively. 
The Newtonian curvature is identical to the Newtonian potential with the
opposite sign ($\psi=-~\phi$)
in the matter-dominated era, where there is vanishing 
anisotropic stress.
The conservation of energy momentum provides the continuity and Euler 
equations,
\bear
\dot\delta+{k\over a}~v&=&-~3~\dot\phi, \\
\dot v+Hv&=&{k\over a}~\psi,
\enar
where $v$ is the velocity of non-relativistic matter in units of $c$.
In the conformal Newtonian gauge, the relativistic equations on sub-horizon
scales correspond to the usual Newtonian equations,
\bear
\label{eq:poisson}
\nabla^2\psi&=&{3H_0^2\over2}~\OM~{\delta\over a},\\
\label{eq:peculiar}
\bdv{v}&=&-~{2\over3}~{a^2Hf\over\OM H_0^2}~\nabla\psi,
\enar
where $f=d\ln D/d\ln a$ and $D$ is a growth factor of the matter density 
perturbation. The evolution of the density perturbation is related to the 
Newtonian potential by
\beeq
\ddot \delta+2H\dot\delta=\nabla^2\psi,
\eneq
and the growth factor $D$ is a growing solution of this differential equation,
normalized to a unity at $a_0$.
Full relativistic consideration results in additional multiple terms in the
right-hand side of the equation (e.g., \citep{MUFEBR92}),
but they are suppressed at least by the
ratio of a characteristic scale $1/k$ to the Hubble distance $1/H$.
Note that we have interchangeably expressed equations in Fourier space and
configuration space, which is valid to the linear order in perturbations
and significantly simplifies the manipulations.

\subsection{Geodesic Equations and Sachs-Wolfe Effects}
\label{ssec:sachs}
The propagation of light rays is described by a photon geodesic 
$x^a(\lambda)$ with an affine parameter $\lambda$, and a null vector 
$k^a=dx^a/d\lambda$ tangent to $x^a$ is determined by the null equation 
($ds^2=k^ak_a=0$) and the geodesic equations ($k^a{_{;b}}k^b=0$). 
In a perturbed FLRW universe, the null vector can be expressed as
\beeq
k^0={\nu\over a}~(1+\delta\nu),~~~
k^\alpha=-~{\nu\over a}~(e^\alpha+\delta e^\alpha),
\label{eq:null}
\eneq
where $\nu$ and $e^\alpha$ are the photon frequency and its (time-reversed)
propagation direction from the observer, and the dimensionless quantities
$\delta\nu$ and 
$\delta e^\alpha$ represent their perturbations.
In a homogeneous expanding universe,
the null vector follows the usual relations $\nu\propto1/a$,
$e^\alpha e_\alpha=1$, and $de^\alpha/d\eta=e^\beta e^\alpha{_{|\beta}}$, and
indeed Eq.~(\ref{eq:null}) may be derived from the null and 
the geodesic equations.
For a comoving observer whose rest frame has vanishing energy flux,
the four velocity is $u^a=(1/a,0)$ and the observed frequency $\nu_\up{obs}$
of a photon source is related to the frequency $\nu_\up{e}$ at the emission
by a redshift parameter,
\beeq
1+z={(k^a~u_a)_\up{e}\over(k^a~u_a)_\up{obs}}={\nu_\up{e}\over\nu_\up{obs}}
={1\over a_\up{e}},
\label{eq:redshift}
\eneq
where we assumed $a_\up{obs}=a_0=1$.

In an inhomogeneous universe, the observed redshift $\zobs$ deviates from 
the true redshift $z$. Perturbations in the null equation is
\beeq
e^\alpha~\delta e_\alpha=\delta\nu+\psi-\phi,
\eneq
and perturbations in the geodesic equations for the temporal 
and spatial components are
\bear
{d\over dy}(\delta\nu+\psi)&=&\psi_{,\alpha}~e^\alpha-{d\phi\over d\eta},
\label{eq:zerogeo}\\
{d\over dy}(\delta e^\alpha+2\phi ~e^\alpha)&=&\delta e^\beta~ e^\alpha{_{|\beta}}
-\delta\nu~{de^\alpha\over d\eta}+\psi^{|\alpha}-\phi^{|\alpha},~~~~~
\label{eq:ageo}
\enar
where we used the zeroth order null geodesic
$d/dy\equiv\partial_\eta-e^\alpha\partial_\alpha=(a/\nu)(d/d\lambda)$
and kept the terms to the first order in perturbations.

The four velocity of a comoving observer is now
$u^a=((1-\psi)/a,~v^\alpha/a)$ and the observed redshift is 
\beeq
1+\zobs={(k^a~u_a)_\up{e}\over(k^a~u_a)_\up{obs}}
=(1+z)\left[1+(\delta\nu+\psi+v_\alpha~ e^\alpha)^\up{e}_\up{o}\right].
\eneq
This can be further simplified by using Eq.~(\ref{eq:zerogeo}) as
\bear
\label{eq:gSW}
1+\zobs&=&(1+z)\times\bigg[1+V(z)-V(0)\\
&-&\psi(z)+\psi(0)+\int_0^y dy~{\partial\over\partial\eta}
(\phi-\psi)\bigg],\nonumber
\enar
where $V=v_\alpha ~e^\alpha$ is the line-of-sight velocity
 \citep{SAWO67,HWNO99,MATSU00}.  The additional terms in the square
bracket alter the simple redshift-distance relation in 
Eq.~(\ref{eq:redshift}), giving rise to the standard redshift-space distortion
by peculiar velocities, the Sachs-Wolfe effect by gravitational redshift, and
the integrated Sachs-Wolfe effect by the time evolution of gravitational 
potential across which photons propagate.
Hereafter
we will collectively refer to these effects as the generalized Sachs-Wolfe
effect.

\subsection{Gravitational Lensing}
\label{ssec:lensing}
In a homogeneous universe, the gravitational lensing effects vanish
and light rays propagate with the direction unchanged.
For a photon source at $\bdv{\hat z}$-axis in an inhomogeneous universe,
the propagation direction from the observer is
$\Vang=e^\alpha=(0,0,1)$ and the null vector is
$k^{x,y}=-(\nu/a)\delta e^{x,y}$. The null vector is further
related to the photon position $r(\chi)\Vang=(x,y)$ on the
sky at any time by
\beeq
k^{x,y}={d\over d\lambda}(r\Vang)={\nu\over a}{d\over dy}(r\Vang),
\eneq
where we replace the derivative with respect to the affine parameter by using
the zeroth order null geodesic, consistent to the first order in perturbations.
The spatial component of the geodesic equation (Eq.[\ref{eq:ageo}]) is then
\beeq
{d\over dy}(\delta e^{x,y})=-{d^2\over dy^2}(r\Vang)=
\psi^{|\alpha}-\phi^{|\alpha}=2~\psi^{|\alpha}.
\label{eq:path}
\eneq

Since gravitational lensing conserves the surface brightness, the observed
surface brightness $I_\up{obs}(\Vang)$ on the sky is simply
the intrinsic surface 
brightness at the source position $\Sang$: $I_\up{obs}(\Vang)=I(\Sang)$,
and the source
position $\Sang$ can be obtained by integrating Eq.~(\ref{eq:path})
along the photon geodesic
\beeq
\Sang=\Vang+\ang~\Psi(\Vang),
\label{eq:lens}
\eneq
with the projected lensing potential
\bear
\Psi(\Vang)&=&-~2\int_0^{y_s}dy'\int_0^{y'}dy ~{\psi(y)\over 
r(\chi_s)~r(\chi)} \nonumber \\
&=&-~2\int_0^{y_s}dy~\psi(y)~{r(\chi_s-\chi)\over r(\chi_s)~r(\chi)},
\enar
where $\ang$ is the derivative with respect to $\Vang$, and
$\chi_s=\int_0^{z_s}dz/H(z)$ is the comoving line-of-sight distance to the
source redshift $z_s$.
The integration along the unperturbed photon geodesic $dy$ is often called
the Born approximation. Following the literature, we take the geodesic
as the photon radial direction $d\chi$, but note that 
$d/d\chi=\partial_\eta-\partial_\chi$.

The convergence $\kappa(\Vang)$ is defined as
$\ang^2\Psi(\Vang)=-2\kappa(\Vang)$ and it is further related
to density fluctuations along the geodesic by Poisson's equation
(Eq.[\ref{eq:poisson}])
\bear
\kappa(\Vang)&=&\int_0^{\chi_s}d\chi~(\nabla^2-\nabla^2_\chi)~
\psi[r(\chi)\Vang,\chi]~
{r(\chi_s-\chi)~r(\chi)\over r(\chi_s)}~~~ \nonumber \\
&=&{3H_0^2\over2}\OM\int_0^{\chi_s}\!\!d\chi~
{\delta[r(\chi)\Vang,\chi]\over a(\chi)}
~{r(\chi_s-\chi)~r(\chi)\over r(\chi_s)}.
\label{eq:conv}
\enar
The contribution from the radial derivatives $\nabla^2_\chi$
is proportional to the potential
difference between the source and observer, and this boundary term is 
negligible compared to the first term \citep{JASEWH00,HISE03a}.
Numerical ray tracing experiments through 
$N$-body simulations show that the weak lensing approximation to the first
order in perturbations is accurate even in nonlinear regime when nonlinear
matter power spectrum is used in place of linear matter power spectrum
\citep{JASEWH00}. Also note that all the prior results for a single source
redshift can be readily generalized to a source population with a redshift
distribution $W(\chi_s)$ by integrating the results over $\chi_s$ with 
$W(\chi_s)$ in the integrand.

While conservation of surface brightness guarantees that photons are neither
destroyed nor created, gravitational deflection
distorts the cross-section of a bundle of light rays, magnifying (or 
de-magnifying) observed fluxes. Gravitational lensing
magnification $\mu(\Vang)$ 
is related to the Jacobian of a mapping from the image plane to the source
plane by
\bear
\mu(\Vang)^{-1}&=&\left|{d^2\Sang\over d^2\Vang}\right|
=\left|\bdv{I}+\ang\ang~\Psi(\Vang)\right| \\
&=& \left|\left[1-\kappa(\Vang)\right]^2-\gamma^2(\Vang)\right|, \nonumber
\enar
where $\bdv{I}$ is a unit $2\times2$ matrix and
$\gamma(\Vang)$ is the tangential shear. In the weak lensing regime,
$\mu(\Vang)=1+2~\kappa(\Vang)$.

Gravitational lensing also modifies the propagation time of light rays in two
ways, compared to the light travel time in the absence of the gravitational
lensing effects: it distorts the photon geodesic, increasing the path length
that photons travel, and the gravitational potential retards the light
travel time. The former is referred to as the geometric time delay
\citep{BLNA86}
\beeq
\tau_\up{geo}(\Vang)={1\over2}~{r(\chi_l)~r(\chi_s)\over r(\chi_s-\chi_l)}~
\ang\Psi(\Vang)\cdot\ang\Psi(\Vang),
\eneq
and the latter is the potential or Shapiro time delay \citep{SHAPIRO64}
\beeq
\tau_\up{pot}(\Vang)={r(\chi_l)~r(\chi_s)\over r(\chi_s-\chi_l)}~
\Psi(\Vang).
\eneq
These effects can be derived by using the small angle approximation in 
deflection and the relation $d\eta=(1-\psi+\phi)d\chi$ from the metric
in Eq.~(\ref{eq:metric}).
Note that the proper time delay can be obtained by multiplying the lens
redshift $1+z_l$ in the limit of a single lens case, and this derivation in
a cosmological context recovers the standard relation for time delay.

\section{Source Fluctuations}
\label{sec:density}
Inhomogeneous matter fluctuations in the universe deflect the propagation
of light rays, giving rise to the gravitational lensing effects.
The generalized Sachs-Wolfe effect also 
arises from the same matter fluctuations responsible for the gravitational
lensing effects. Having discussed the basic mechanism of the
gravitational lensing and the generalized Sachs-Wolfe effects
that complicate the simple interpretation of observable quantities,
we now investigate their impact on an observed overdensity field 
$\dobs(\Vang,z)$ of source galaxies. Contributions to $\dobs(\Vang,z)$
come from matter fluctuations in addition 
to the intrinsic overdensity $\delta(\Vang,z)$ of source galaxies.
Noting that the contributions can be linearized and added
to the first order in perturbations, we separate these contributions
as two physically distinct parts: one that involves the 
change of volume, and one that involves the intrinsic properties 
of source galaxies.
The impact on galaxy two-point statistics will be discussed in 
the following section.

\subsection{Volume Effect}
\label{ssec:vol}
Consider a unit comoving volume $dV=r^2(\chi)d\Omega dz/H(z)$
and a unit flux interval $df$, and let
$n(\Vang,z,f)$ be the comoving number density of source galaxies.
The generalized Sachs-Wolfe effect alters the unit comoving volume $dV$.
Note, however, that it not only changes the unit redshift interval $dz$, 
but also changes both the angular diameter distance $r(\chi)$ and the
Hubble parameter $H(z)$. By imposing the number conservation,
the observed number density of the source galaxies can be obtained by
\beeq
\label{eq:nSW}
\nobs(\zobs)=n(z)\left[1+\delta V\right]
{r^2(\chi)\over r^2(\chi_\up{obs})}
{H(\zobs)\over H(z)}{dz\over d\zobs},
\eneq
where $\delta V=(2\phi+\varepsilon)^e_o$ represents the distortion of volume
element, when it is transformed from the conformal Newtonian gauge to
the local Lorentz frame, where the velocity of non-relativistic matter
vanishes. We give a more rigorous derivation in Appendix~\ref{app:ngal}.
The solid angle $d\Omega$ remains unaffected by the generalized Sachs-Wolfe
effect.

If the mean comoving number density evolves slowly compared
to the redshift change due to the generalized Sachs-Wolfe effect
$\bar n(z)=\bar n(\zobs)$,
contributions to $\dobs(\zobs)$ arise solely from the change in
volume element $dV$,
\beeq
\label{eq:dgSW}
\dobs(\zobs)=\delta(z)-2~{1+z\over H\chi}~\varepsilon  
-(1+z)H~{d\over dz}\left({\varepsilon\over H}\right) 
-\varepsilon+\delta V,
\eneq
where we rewrote Eq.~(\ref{eq:gSW}) as $1+\zobs=(1+z)(1+\varepsilon)$ and
the contribution $\varepsilon$
from the generalized Sachs-Wolfe effect is
\beeq
\varepsilon(z)= V(z)-V(0)-\psi(z)+\psi(0)-2
\int_0^\chi d\chi{\partial\psi\over\partial\eta}.
\label{eq:varep}
\eneq
In the Einstein-de~Sitter universe, 
the Newtonian potential is constant and hence the
integrated Sachs-Wolfe effect vanishes. In general, as we show in the next
section, the peculiar velocity effect is dominant over the
Sachs-Wolfe and the integrated Sachs-Wolfe effects, and
$\varepsilon(z)\simeq V(z)-V(0)$. 
Note that while our derivation so far is valid for nonflat universes,
in deriving Eq.~(\ref{eq:dgSW}) we assumed that the spatial curvature $K$
is close to zero. The second term in Eq.~(\ref{eq:dgSW}) has a multiplicative
factor $\sqrt{K}\chi/\tan(\sqrt{K}\chi)$ for a closed universe $K>0$ and
$\sqrt{-K}\chi/\tanh(\sqrt{-K}\chi)$ for an open universe $K<0$, which
becomes a unity as $K\rightarrow0$.

With a proper line-of-sight distance $r_p=\chi(z)/(1+z)$ and a
normalized peculiar velocity
$u=V(z)/H(z)$, Eq.~(\ref{eq:dgSW}) can be rearranged as
\beeq
\dobs(\zobs)=\delta(z)-{2u\over r_p}-{du\over dr_p},
\label{eq:stdz}
\eneq
if we ignore the Sachs-Wolfe and the integrated Sachs-Wolfe effects in
Eq.~(\ref{eq:varep}).
This recovers the standard relation for redshift-space distortions
\citep{KAISE87,STWI95,HAMIL98}. Note that
the standard method ignores the contributions
in Eq.~(\ref{eq:dgSW}) from the Sachs-Wolfe and the integrated Sachs-Wolfe
effects. We discuss their impact in Sec.~\ref{sec:two}.

Gravitational lensing magnification increases
the flux interval $df$ and the solid angle $d\Omega$ by a factor of
$\mu$, respectively. With the number conservation in $dV$ and $df$,
the observed number density is therefore
\beeq
\nobs(\fobs)=n(f)~{df\over d\fobs}~{d\Omega\over d\Omega_\up{obs}}=
{1\over\mu^2}~n(f).
\label{eq:mag}
\eneq
Similarly, if the mean comoving number density is the same over the 
flux change due to lensing magnification (i.e., the source
luminosity function is flat), the observed overdensity is then
\beeq
\dobs(\Vang)=\delta(\Vang)-4~\kappa(\Vang),
\eneq
reflecting the change in volume and flux.

Gravitational lensing displaces the source position on the sky 
according to Eq.~(\ref{eq:lens}), and the observed number density is 
$\nobs(\Vang)=n\left[\Vang+\ang\Psi(\Vang)\right]$. By Taylor expanding
$\nobs(\Vang)$ to the first order in $\Psi(\Vang)$,
the observed overdensity can be written as
\beeq
\dobs(\Vang)=\delta(\Vang)+\ang\Psi(\Vang)\cdot\ang\delta(\Vang).
\label{eq:gld}
\eneq
Note that the additional contribution is already in the second order in
perturbations and furthermore it vanishes on average, because the deflection
angle $\ang\Psi(\Vang)$ has no preferred direction. The first non-vanishing
effect from gravitational
lensing displacement comes in the second order in $\Psi(\Vang)$
\citep{VADOET07}, and we therefore ignore this effect.

Finally the gravitational time delay decreases the arrival time of photons
in an overdense region, compared to that in the absence of
lensing. The net effect is therefore that we sample sources at farther 
distance in the fixed time interval \citep{HUCO01}.
However, for discrete sources
the effect vanishes as long as the life time of the sources is longer
than the time delay.

\subsection{Source Effect}
\label{ssec:src}
The generalized Sachs-Wolfe and the gravitational lensing effects modify a
unit volume and a unit flux interval, leading to the contributions to
$\dobs(\Vang,z,f)$. Furthermore, the changes in observed redshift and flux
can result in different mean number densities, if the redshift distribution of
the source galaxy population varies
in the redshift interval or the luminosity function is non-trivial
over the flux change. These additional contributions from the change
in mean number densities are related to the intrinsic properties of source 
galaxies, and we collectively refer to these effects as the source effect.
However, note that while the source effect may be absent for some galaxy
populations, the volume effect is always present. Therefore, we keep together
the contributions from the volume effect in considering the source effect.

We first consider the effect of gravitational lensing magnification.
Lensing magnification not
only increases $d\Omega$ and $df$ in Eq.~(\ref{eq:mag}), but also changes
the number count of source galaxies, 
if the luminosity function is non-flat, i.e.,
$\bar\nobs(\fobs)\ne\bar n(f)$. Assuming $\bar n(f)df\propto f^{-s}df$
with a constant slope $s$ over a narrow flux range $df$, the observed
number density can be expressed as
\beeq
\bar\nobs(\fobs)={\bar n(\fobs/\mu)\over\mu^2}=\bar n(\fobs)~\mu^{s-2},
\eneq
and the observed overdensity is now
\bear
\label{eq:mgb}
\dobs(\Vang)&=&\delta(\Vang)+(2s-4)\kappa(\Vang) \\
&=&\delta(\Vang)+5(p-0.4)\kappa(\Vang), \nonumber 
\enar
where we used the logarithmic slope $p=d\log\bar n(m)/dm=0.4(s-1)$ 
in a sample with limiting magnitude $m$. In the literature,
these contributions from both the volume and the source effects
are referred to as the magnification bias 
\citep{NARAY89,BARTE95,JAIN02,JASCSH03,MOJA98}. Note that this bias can be 
either positive or negative, depending on the slope $p$, and the volume
effect can be canceled by the source effect with $p=0.4$
(see \citep{SCMEET05} for the recent detection from the Sloan Digital
Sky Survey).

The redshift distribution of source galaxies 
also affects the mean number counts due
to the generalized Sachs-Wolfe effect. For a redshift distribution 
$\bar n(z)dz\propto z^\alpha\exp\left[-(z/z_0)^\beta\right]dz$,
the observed overdensity can be obtained by substituting $\bar n(z)$ 
with $\bar n\left[\zobs-(1+\zobs)\varepsilon\right]$,
\bear
\label{eq:zevo}
\dobs(z)&=&\delta(z)-{1+z\over z}\left[\alpha-\beta
\left({z\over z_0}\right)^\beta\right]\varepsilon \\
&-&2~{1+z\over H\chi}~\varepsilon
-(1+z)H{d\over dz}
\left({\varepsilon\over H}\right)-\varepsilon+\delta V,  \nonumber
\enar
where the second term in the right-hand side is
the additional contribution related to the evolution of source galaxies, and
the rest of the additional terms come from the volume effect in 
Eq.~(\ref{eq:dgSW}).

\subsection{Summary}
\label{ssec:sum}
We have investigated the effects of inhomogeneous matter fluctuations on 
observed overdensity fields. Here we summarize their contributions and 
clarify the functional dependence. We then compare their impact on galaxy
two-point statistics in Sec.~\ref{sec:two}.

For a sample of galaxies at redshift $z$ selected with a limiting flux $f$ and
narrow intervals of $dz$ and $df$, the observed overdensity 
$\dobs(\Vang,z,f)$ is the sum of
the intrinsic overdensity field $\delta(\Vang,z,f)$
and the contributions from the gravitational 
lensing and the generalized Sachs-Wolfe effects:
\beeq
\dobs(\Vang,z,f)=\delta+\dMB+\zdist+\devo.
\label{eq:summary}
\eneq
From Eq.~(\ref{eq:mgb}), the magnification bias is defined as
\beeq
\dMB(\Vang,z,f)=5\left[p(f)-0.4\right]\kappa(\Vang,z),
\eneq
with redshift $z$ being the source redshift of the convergence $\kappa(\Vang)$
in Eq.~(\ref{eq:conv}). Considering $\varepsilon(z)\simeq V(z)-V(0)$, we
call the volume effect in Eq.~(\ref{eq:dgSW}) as 
the redshift-space distortion bias,
\bear
\label{eq:zfive}
\zdist(\Vang,z)&=&-2~{1+z\over H\chi}~\varepsilon
-(1+z)H{d\over dz}
\left({\varepsilon\over H}\right)-\varepsilon+\delta V \nonumber \\
&=&-2~{1+z\over H\chi}~\varepsilon +{1+z\over H}~\varepsilon~{dH\over dz} 
\nonumber \\
&&-{1+z\over H}~{\partial\varepsilon\over\partial\chi} 
-\varepsilon+\delta V. 
\enar
Note that the generalized Sachs-Wolfe effect $\varepsilon(z)$ implicitly
depends on the direction $\Vang$ via the line-of-sight velocity
$V(z)=v_\alpha e^\alpha=\Vang\cdot\bdv{v}(\Vang,z)$, but it is independent
of the limiting flux $f$, provided that galaxies have no velocity bias
(i.e., galaxies and matter follow the same velocity field).
Finally, the evolution bias is defined from Eq.~(\ref{eq:zevo}) as
\beeq
\devo(\Vang,z,f)=-{1+z\over z}\left[\alpha-\beta
\left({z\over z_0}\right)^\beta\right]\varepsilon,
\label{eq:bevo}
\eneq
where the directional dependence comes from $\varepsilon$ and the evolution
coefficients $(\alpha,\beta,z_0)$ depend on the galaxy sample selected with
the limiting flux $f$. While the evolution bias arising from the difference 
between $\bar n(z)$ and $\bar n(\zobs)$ was recognized 
\citep{KAISE87,HAMIL98,MATSU04},
it has been ignored in the literature. 
However, we show in Sec.~\ref{sec:two} that the evolution bias
can be significantly enhanced. Last, we want to emphasize that 
equation~(\ref{eq:summary}) is gauge-invariant as is
written in the conformal Newtonian gauge.

\section{Galaxy Two-Point Statistics}
\label{sec:two}
We have derived additional contributions of the gravitational lensing and 
the generalized Sachs-Wolfe effects to the intrinsic density fluctuations in
Sec.~\ref{sec:density}, fully consistent up to the first order
in perturbations. Given two samples of galaxies with limiting fluxes
$f_1$ and $f_2$, the observed galaxy correlation function is then
$\xiobs(\Vang_1,z_1,\Vang_2,z_2)=\langle\dobs(\Vang_1,z_1)~\dobs(\Vang_2,z_2)
\rangle$ and the observed power spectrum is 
$\langle\dobs(\bdv{k}_1,z_1)~\dobs^*(\bdv{k}_2,z_2)\rangle=(2\pi)^3
\delta^D(\bdv{k}_1-\bdv{k}_2)P_\up{obs}(k_1)$.
In Sec.~\ref{ssec:two},
we derive this ensemble average of all the combinations of each component in 
$\dobs$ in Eq.~(\ref{eq:summary}), after we simplify the equation.
We then discuss their impact on the observed galaxy two-point statistics
by analyzing specific examples in Sec.~\ref{ssec:com}.

\subsection{Correlation Function and Power Spectrum}
\label{ssec:two}
Here we compute the observed galaxy correlation function $\xiobs$ and
power spectrum $P_\up{obs}$. However, as some components in $\dobs$ are 
smaller than other components, their combinations are even smaller 
by an order-of-magnitude. We therefore
start by estimating the auto-correlation functions of each component 
and simplify the equation before we compute all the
cross-correlation functions and power spectra.

We first consider the correlation of the redshift-space distortion bias
$\xizz=\langle\zdist(\Vang_1,z_1)~\zdist(\Vang_2,z_2)
\rangle$. The redshift-space distortion bias $\zdist$ in Eq.~(\ref{eq:zfive})
has five components that depend either $\varepsilon(z)$ or its partial 
derivative with respect to $z$ or $\chi$, and the contribution 
$\varepsilon(z)$ from the generalized Sachs-Wolfe effect in 
Eq.~(\ref{eq:varep}) has also three different components that depend on the
peculiar velocity, the Newtonian potential, and its time derivative.
The Newtonian potential and the peculiar velocity in 
Eqs.~(\ref{eq:poisson}) and~(\ref{eq:peculiar}) take the simple form in
Fourier space
\bear
\psi_\bdv{k}&=&-{3H_0^2\over2}~{\OM\over a}~{\delta_\bdv{k}\over k^2},\\
\bdv{v}_\bdv{k}&=&iHfa~\delta_\bdv{k}~{\bdv{k}\over k^2}.
\enar
On a typical correlation scale $1/k$, they scale as
$H^2\delta_k/k^2$ and $H\delta_k/k$ with $f\simeq1$ at $z\gtrsim1$:
$\psi_k$ is smaller than $v_k$ by the ratio of the correlation scale $1/k$
to the Hubble distance $1/H$. Similarly, the integrated Sachs-Wolfe effect
is of the same order as the Newtonian potential and it vanishes in the limit
of zero cosmological constant, i.e., Einstein-de~Sitter universe,
because it is proportional to the time derivative of the ratio of the growth 
factor to the expansion scale factor $D(z)/a$.
Therefore, we can safely ignore the Sachs-Wolfe and the integrated
Sachs-Wolfe effects and we assume $\varepsilon(z)\simeq V(z)$. Note that
given a particular realization of the observer's rest frame, 
its peculiar velocity $V(0)$ is uncorrelated and the unobservable potential 
$\psi(0)$ in Eq.~(\ref{eq:varep})
can be absorbed by a gauge transformation.

With the assumption $\varepsilon(z)\simeq V(z)$, we further simplify
Eq.~(\ref{eq:zfive}) by comparing the five components in the redshift-space 
distortion bias, and similar justification was made in \citep{MATSU00}.
Respectively, each component scales as 
$\delta_k/k\chi$, $H\delta_k/k$,
$\partial\delta_k/k\partial\chi$, $H\delta_k/k$, 
and $H\delta_k/k$, and hence
they are smaller than $\delta_k$ by the ratio of
correlation scale $1/k$ to the Hubble distance $1/H$ or the 
line-of-sight distance $\chi$ (roughly of order $1/H$), except the third
component: the partial derivative with respect to $\chi$ cancels 
the correlation scale $1/k$ and hence the amplitude of the third component
is of order $\delta_k$, larger than the other components in the redshift-space
distortion bias. Therefore, we only keep the dominant component in
the redshift-space distortion bias \citep{MATSU00},
\beeq
\zdist(\Vang,z)\simeq-{1+z\over H}~{\partial V\over\partial\chi},
\eneq
consistent with the standard relation for the redshift-space distortion,
justifying its nomenclature. However, note that all these ignored components
are proportional to $\varepsilon$. At low redshift, they contribute 
to galaxy two-point statistics at the sub-percent level, while we show 
in Sec.~\ref{ssec:com} that at higher 
redshift their contribution is somewhat larger.

Having substantially reduced the number of combinations for an ensemble 
average, we are now well positioned to compute correlation functions and
their power spectra.
For two galaxy positions $\bdv{x}_1=\left[r(\chi_1)\Vang_1,\chi_1\right]$
and $\bdv{x}_2=\left[r(\chi_2)\Vang_2,\chi_2\right]$, the auto-correlation 
of the redshift-space distortion bias is
\bear
\label{eq:pzz}
\xizz&=&\langle\zdist(\Vang_1,z_1)~\zdist(\Vang_2,z_2)\rangle\\
&=&f_1f_2\int{d^3\bdv{k}\over(2\pi)^3}~e^{i\bdv{k}\cdot(\bdv{x_1-x_2})}
\Pm(k;z_1,z_2)~{k_z^4\over k^4} \nonumber \\
&=&f_1f_2\int_0^\infty{dk\over k}~{k^3\over 2\pi^2}~\Pm(k;z_1,z_2) \nonumber \\
&\times&
\left[{1\over5}j_0(k\rtd)P_0(\gamma)-{4\over7}j_2(k\rtd)P_2(\gamma)+
{8\over35}j_4(k\rtd)P_4(\gamma)\right], \nonumber
\enar
where the 3D comoving separation is 
$\rtd=\left[r(\bar\chi)^2\Delta\theta^2+(\chi_2-\chi_1)^2\right]^{1/2}$ with
$\Delta\theta=|\Vang_1-\Vang_2|$ and
$\bar\chi=(\chi_1+\chi_2)/2$, and the angle subtended by the comoving 
separation is
$\gamma=\cos\Delta\theta=(\chi_2-\chi_1)/\rtd$. $P_n(x)$ and $j_n(x)$ 
are the $n$-th order Legendre polynomial and spherical Bessel function,
respectively. We assumed
the distant observer approximation such that $k_z$ is the line-of-sight 
component of the wavenumber $k$, but it can be relaxed by replacing
$k_z^4$ by $\left[(\Vang_1\cdot\bdv{k})(\Vang_2\cdot\bdv{k})\right]^2$.
The linear matter power spectrum 
is computed by $\Pm(k;z_1,z_2)=D(z_1)D(z_2)\Pm(k)$, while
we use $\Pm(k;z_1,z_2)=\Pm(k;\bar z)$ with $\bar z=(z_1+z_2)/2$ when we
compute the effect of the nonlinear matter power spectrum using the
\citet{SMPEET03} approximation. The power spectrum of the redshift-space
distortion bias can be readily read off
from Eq.~(\ref{eq:pzz}) and its power is boosted along the line-of-sight by
$f_1f_2\mu_k^4$ with $\mu_k=k_z/k$.

Next we consider the correlation of the evolution bias. The observed redshift
$\zobs$ is different from the true redshift $z$ due to the generalized
Sachs-Wolfe effect and the redshift distribution 
of the source mean number density 
gives rise to the evolution bias. The evolution bias $\devo$ is proportional
to $\varepsilon(z)\simeq V(z)$ and it is typically smaller than $\zdist$ by
the ratio of a correlation scale $1/k$ to the Hubble distance $1/H$. However,
beyond the mean redshift of source populations, the mean number density 
changes exponentially and the evolution bias can be substantially boosted by
the prefactor 
\beeq
E(z;f)=-{1+z\over z}\left[\alpha-\beta\left({z\over z_0}\right)^\beta\right],
\label{eq:eboo}
\eneq
defined such that Eq.~(\ref{eq:bevo}) becomes $\devo=E(z)~\varepsilon(z)$.
While the exact functional form of $E(z)$ depends on the assumed redshift 
distribution, it captures the general trend of the enhancement in $\devo$
beyond the mean redshift. The correlation of the evolution bias is therefore
\bear
\xievo&=&\langle\devo(\Vang_1,z_1)~\devo(\Vang_2,z_2)\rangle\\
&=&(HfaE)_1(HfaE)_2\!\!\int\!\!\!
{d^3\bdv{k}\over(2\pi)^3}e^{i\bdv{k}\cdot(\bdv{x_1-x_2})} 
\Pm(k;z_1,z_2){k_z^2\over k^4} \nonumber \\
&=&(HfaE)_1(HfaE)_2\int_0^\infty{dk\over k}{k\over 2\pi^2}~\Pm(k;z_1,z_2) 
\nonumber \\
&\times&\left[{1\over3}j_0(k\rtd)P_0(\gamma)-{2\over3}j_2(k\rtd)
P_2(\gamma)\right], 
\nonumber
\enar
where the subscripts in the round brackets represent that the products
$(HfaE)$ are
computed at $z_1$ and $z_2$. Its power spectrum is also anisotropic and has
structure similar to the redshift-space distortion bias.

Finally, the inhomogeneous matter fluctuations along the two lines-of-sight 
result in the correlation of the magnification bias
\bear
\label{eq:xiMB}
\xiMB&=&\langle\dMB(\Vang_1,z_1)~\dMB(\Vang_2,z_2)\rangle  \\
&=&(5p_1-2)(5p_2-2)\left({3H_0^2\over2}\OM\right)^2  
\int_0^{\chi_1}d\chi\left[{r(\chi)\over a(\chi)}\right]^2 \nonumber \\
&\times&{r(\chi_1-\chi)\over r_1}{r(\chi_2-\chi)\over r_2}~
w_p\left[r(\chi)\Delta\theta;z\right], \nonumber
\enar
where we used the Limber approximation \citep{LIMBE54} (see 
Appendix~\ref{app:limber}). Without loss of generality,
we assumed $z_1\leq z_2$. The projected correlation function
$w_p(R)$ is obtained by integrating the 3D matter correlation function 
$\xim(x)$ along the line-of-sight at a fixed redshift $z(\chi)$ and 2D
transverse separation $R$,
\bear
w_p\left[R;z\right]&=&\int_{-\infty}^\infty dr_\parallel~
\xim\left[\rtd=\sqrt{R^2+r_\parallel^2};z\right] \\
&=&\int_0^\infty{k~dk\over2\pi}\Pm(k;z)J_0(kR),\nonumber
\enar
where $J_n(x)$ is the $n$-th order Bessel function of the first kind.
Assuming that the source redshifts are sufficiently high and hence $\xiMB$
is independent of $z_1$ and $z_2$, the power spectrum of the magnification 
bias is 
$P_\up{mb}=(2\pi)\delta^D(k_z)(5p_1-2)(5p_2-2)r^2(\bar\chi)
C^\up{\kappa\kappa}_{l=k_\perp r(\bar\chi)}$,
where the angular power spectrum of the convergence is
\beeq
\label{eq:akk}
\clcor{\up{\kappa\kappa}}=\left({3H_0^2\over2}\OM\right)^2 
\int_0^{\bar\chi} d\chi~\left[{r(\bar\chi-\chi)\over a(\chi)~r(\bar\chi)}
\right]^2\Pm\left[k={l\over r(\chi)};z\right].
\eneq
The Dirac delta function results from our assumption
that $\xiMB$ is a function of transverse direction only, but it can be 
somewhat relaxed
by replacing $(2\pi)\delta^D(k_z)$ by a survey window function 
\citep{HUGALO08}. Note that while we are interested in how the magnification
bias affects the 3D correlation of the intrinsic source fluctuations, the 
magnification bias arises from the matter fluctuations along the line-of-sight
(not at a single redshift plane) and thereby angular correlation function and
its angular
power spectrum are better suited for quantifying its statistics. Indeed,
the correlation function of the magnification bias is identical to the angular
correlation function, 
$\xiMB(\Vang_1,z_1,\Vang_2,z_2)=w_\up{mb}(\Delta\theta;z_1,z_2)$, and we relate
2D angular power spectrum to 3D power spectrum by
$P(k)=(2\pi)\delta^D(k_z)r^2(\bar\chi)C_{l=k_\perp r(\bar\chi)}$
(see Appendix~\ref{app:limber}).

With all the additional contributions of the gravitational lensing and the
generalized Sachs-Wolfe effects in hand, the correlation of the intrinsic
fluctuation of sources is modeled using the linear bias model,
\bear
\label{eq:xint}
\xint&=&\langle\delta(\Vang_1,z_1)~\delta(\Vang_2,z_2)\rangle\\
&=&b_1~b_2
\int{d^3\bdv{k}\over(2\pi)^3}~e^{i\bdv{k}\cdot(\bdv{x_1-x_2})}~\Pm(k;z_1,z_2)
\nonumber \\
&=&b_1~b_2 \int_0^\infty{dk\over k}{k^3\over2\pi^2}~\Pm(k;z_1,z_2)~j_0(k\rtd),
\nonumber
\enar
where the constant linear bias factors $b_1$ and $b_2$ are the ratio of
the intrinsic source fluctuation to the underlying matter fluctuation at
$z_1$ and $z_2$.

To complete our calculations of $\xiobs$,
we now compute the cross-correlation functions and power spectra between
the intrinsic source fluctuation  and the fluctuations from the 
gravitational lensing and the generalized Sachs-Wolfe effects. First, the
redshift-space distortion bias and the intrinsic source 
fluctuation provide two cross-correlation functions
\bear
\label{eq:xidz}
\xi_{\delta\up{z}}&=&\langle\delta(\Vang_1,z_1)~\zdist(\Vang_2,z_2)\rangle\\
&=&b_1~f_2\int{d^3\bdv{k}\over(2\pi)^3}~e^{i\bdv{k}\cdot(\bdv{x_1-x_2})}
~\Pm(k;z_1,z_2)
~{k_z^2\over k^2} \nonumber \\
&=&b_1~f_2\int_0^\infty{dk\over k}~{k^3\over2\pi^2}~\Pm(k;z_1,z_2) \nonumber \\
&\times&\left[{1\over3}j_0(k\rtd)P_0(\gamma)-{2\over3}j_2(k\rtd)
P_2(\gamma)\right], 
\nonumber
\enar
and similarly for
$\xi_{\up{z}\delta}=\langle\zdist(\Vang_1,z_1)~\delta(\Vang_2,z_2)\rangle$
with the two indices exchanged in Eq.~(\ref{eq:xidz}).
Combined with $\xizz$ in Eq.~(\ref{eq:pzz}), these two cross-correlation
functions constitute the standard redshift-space correlation function
\beeq
\xi_\up{z-dist}=\xint+\xizz+\xizd+\xidz
=\sum_{l=0,2,4}P_l(\gamma)~\xi_l(\rtd), 
\label{eq:z-dist}
\eneq
which is often expressed in terms of the multipole components
\citep{HAMIL92,COFIWE94,HAMIL98}
\beeq
\xi_l=c_l(\beta_1,\beta_2)~b_1~b_2~i^l\int_0^\infty{dk\over k}~{k^3\over2\pi^2}
~\Pm(k;z_1,z_2)~j_l(k\rtd),
\label{eq:zmul}
\eneq
with its coefficients
\beeq
\label{eq:mulc}
\left(\begin{array}{c}
c_0\\[4pt]c_2\\[4pt]c_4
\end{array}\right)=
\left(\begin{array}{c}
1+{\beta_1+\beta_2\over3}+{\beta_1~\beta_2\over5}\\[4pt]
{2\over3}(\beta_1+\beta_2)+{4\over7}\beta_1~\beta_2\\[4pt]
{8\over35}\beta_1~\beta_2
\end{array}\right),
\eneq
where $\beta=f/b$. Analogously, the redshift-space power spectrum is
\bear
\label{eq:pzf}
P_\up{z-dist}&=&P_\up{int}+P_\up{z\delta}+P_\up{\delta z}+P_\up{zz}\\
&=&\left[1+(\beta_1+\beta_2)~\mu_k^2+\beta_1~\beta_2~\mu_k^4\right]
P_\up{int}(k) \nonumber \\
&=&\sum_{l=0,2,4}P_l(\mu_k)~P^\up{z}_l(k), \nonumber 
\enar
with the intrinsic source power spectrum $P_\up{int}(k)=b_1~b_2~\Pm(k;z_1,z_2)$,
and its multipole components $P_l^\up{z}(k)$
in Fourier space are related to the multipole components $\xi_l(\rtd)$
in real space as
\beeq
P^\up{z}_l(k)=4\pi~i^l\int_0^\infty dx~x^2~\xi_l(x)~j_l(kx).
\eneq

\begin{figure*}
\centerline{\psfig{file=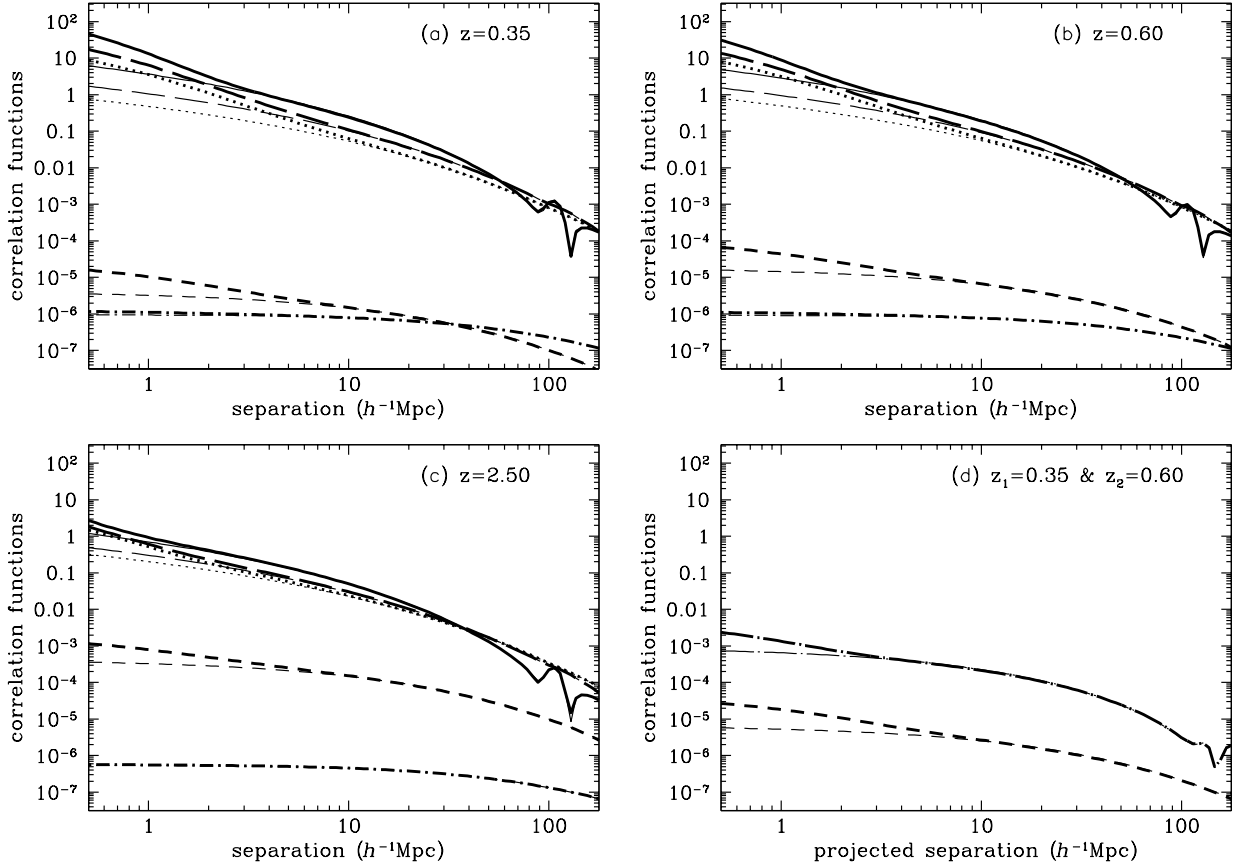, width=7.0in}}
\caption{Dissection of the observed two-point correlation function of galaxies.
Solid, dotted, and long dashed lines represent correlation functions of the
intrinsic source fluctuations $\xint/b^2$ and the redshift-space
distortion bias $\xizz$, and their cross-correlation function 
$\xi_{\delta\up{z}}/b=\xi_{\up{z}\delta}/b$, respectively. Correlation 
functions of the magnification bias $\xiMB/(5p-2)^2$ and the evolution
bias $\xievo/E^2$ are shown as short dashed and short dot-dashed lines.
Note that while
the galaxy bias factor $b$ and magnification bias factor $(5p-2)$ are
of order unity, the evolution boost factor $E$ can be an order of magnitude
larger
(see Fig.~\ref{fig:evo}). The correlation functions are computed by using
the linear ({\it thin}) and the nonlinear ({\it thick}) matter power spectra,
and source galaxies are assumed to be at the same redshift indicated in the
legend. The correlation functions of the intrinsic source fluctuations become 
negative at $\rtd\gtrsim128\hmpc$, where its absolute value is plotted.
Panel~($d$) plots the cross-correlation function 
$\xidMB/b_1(5p_2-2)$ of the intrinsic fluctuation of source
galaxies at $z_1=0.35$ and the magnification bias from source galaxies
at $z_2=0.6$ as long dot-dashed lines.
With large line-of-sight separation $600\hmpc$, only $\xiMB$ and 
$\xidMB$ that depend on projected separation $R$ rather than 3D
separation $\rtd$ itself are appreciable, i.e.,
$\xint\simeq\xi_{\delta\up{z}}\simeq\xi_{\up{z}\delta}\simeq\xizz\simeq
\xievo\simeq0$.}
\label{fig:corr}
\end{figure*}

Since the magnification bias arises from the matter fluctuations along the
line-of-sight, it correlates with the intrinsic source fluctuation at
lower redshift ($z_1<z_2$),
\bear
\label{eq:xidmb}
\xidMB&=&\langle\delta(\Vang_1,z_1)~\dMB(\Vang_2,z_2)\rangle\\
&=&b_1(5p_2-2)\left({3H_0^2\over2}\OM\right)
{r(\chi_2-\chi_1)~r_1\over a_1~r_2}
w_p\left[r_1\Delta\theta;z_1\right], \nonumber
\enar
but the correlation vanishes when the source is at higher redshift, i.e.,
$\xi_{\up{mb}~\delta}=0$. The power spectrum is also related to 
the angular power spectrum of the cross-term
\bear
\label{eq:admb}
\clcor{\delta~\up{mb}}&=&b_1~(5p_2-2)\left({3H_0^2\over2}\OM\right) \\
&\times&{r(\chi_2-\chi_1)\over a_1~r_1~r_2}~
\Pm\left[k_\perp={l\over r_1};z_1\right], \nonumber
\enar
as $P_{\delta~\up{mb}}=(2\pi)\delta^D(k_z)r^2_1
C^{\delta~\up{mb}}_{l=k_\perp r_1}$. 
Finally, all the cross-correlations that involve $\devo$
are zero, since $\devo$ is odd in the line-of-sight component $V$ of peculiar
velocities and the universe has no preferred
direction. Two remaining cross terms $\xi_\up{z~mb}$ and $\xi_\up{mb~z}$
also vanish, since $\zdist$ is proportional to $k_z^2$ and the line-of-sight
fluctuations are smoothed out in the Limber approximation.

\begin{figure}
\centerline{\psfig{file=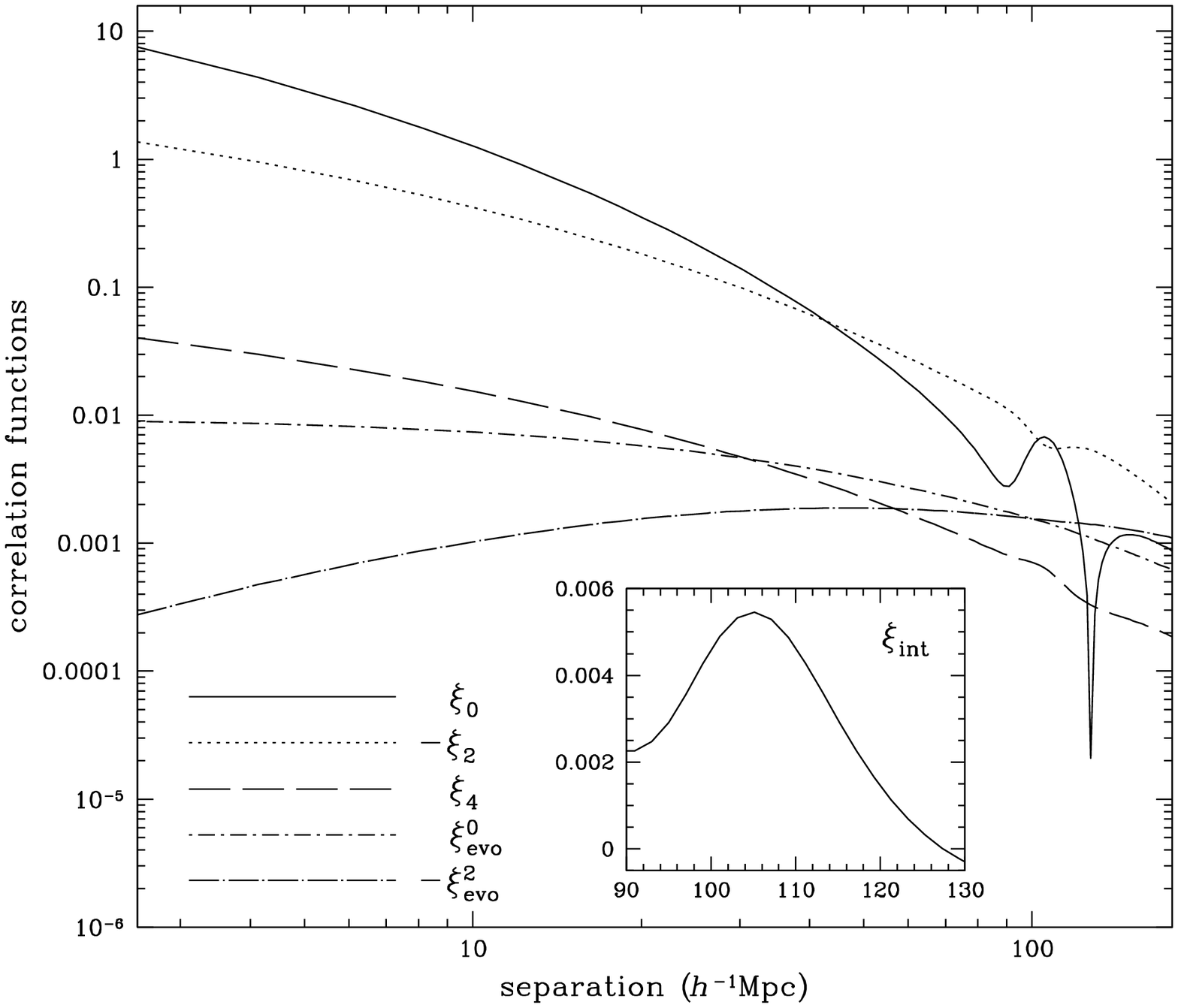, width=3.5in}}
\caption{Multipole components of the redshift-space correlation function
$\xi_\up{z-dist}$ and the evolution bias $\xievo$ at $z=0.35$. 
We define multipole components of $\xievo$ as 
$\xievo=\xievo^0(\rtd)P_0(\gamma)+\xievo^2(\rtd)P_2(\gamma)$, 
in the same way 
multipole components of $\xi_\up{z-dist}$ is defined in Eq.~(\ref{eq:z-dist}).
We assume that the galaxy bias factor is $b=2$ and the evolution boost factor
is $E=100$ for a proper comparison.
The correlation functions are computed by using the linear matter
power spectrum only.
$\xi_0(\rtd)$ becomes negative at $\rtd\gtrsim128\hmpc$, where its
absolute value is plotted. The inset shows the correlation $\xint$
of the intrinsic
source fluctuations around the acoustic scale and it is related to the monopole
by $\xi_0=\xint\cdot c_0$ in Eq.~(\ref{eq:zmul}), where $c_0=1.24$ in
our fiducial model.}
\label{fig:comp}
\end{figure}

\subsection{Comparison}
\label{ssec:com}
To compare the additional contributions to the observed correlation function
$\xiobs$ and power spectrum $P_\up{obs}$,
we consider near-future dark energy surveys that will 
target galaxies and quasars to measure their correlation function and power 
spectrum at high redshifts.
For example, the baryonic oscillation spectroscopic survey (BOSS) will measure
1.5~million luminous red galaxies to determine the angular diameter distances
at $z=0.35$ and~0.6, and use 160,000 quasars to measure the clustering of
Lyman-$\alpha$ forests at $z=2.5$ \citep{WEBEET07,SCBLET07}. 
For the purpose of illustration, we show
our calculations of galaxy two-point statistics at these redshifts.
Here we adopt a flat $\Lambda$CDM universe: the cosmological parameters are
the matter density $\OM h^2=0.137$, the baryon density $\OB h^2=0.0227$, the
Hubble constant $h=0.70$, the spectral index $n_s=0.96$, the optical depth to
the last scattering surface $\tau=0.084$, and the primordial
curvature perturbation amplitude $\Delta_\zeta^2=2.457\times10^{-9}$
at $k=0.002~\mpci$ (corresponding to the
matter power spectrum normalization $\rms=0.817$), consistent with the
recent results (e.g., \citep{TESTET06,SPBEET07,KODUET08}).
The matter transfer function is computed by using
{\scriptsize CMBFAST} \citep{SEZA96}.

Figure~\ref{fig:corr} examines the separate contributions of the gravitational
lensing and the generalized Sachs-Wolfe effects to the observed two-point
correlation function of galaxies. We show the correlation functions of the
intrinsic galaxy fluctuations ($\xint/b^2;~solid$) and the 
redshift-space distortion bias ($\xizz;~dotted$), and their cross-correlation
function ($\xi_{\delta\up{z}}/b=\xi_{\up{z}\delta}/b;~long~dashed$). 
The correlation functions are computed by 
using the linear ($thin$) and the nonlinear
($thick$) matter power spectrum. Note that they only differ on small scales
and the nonlinear effect decreases at high redshift as shown in
Fig.~\ref{fig:corr}$a$ to Fig.~\ref{fig:corr}$c$, going from $z=0.35$ to
$z=2.5$. The source galaxies are assumed to be at the same redshift
$(z=z_1=z_2)$ shown in the figure legend, and thus 3D separation $\rtd$ is
equal to 2D projected separation $R=r(\bar\chi)\Delta\theta$. However,
two galaxy populations are separately placed at $z_1=0.35$ and $z_2=0.6$
in Fig.~\ref{fig:corr}$d$, and the $x$-axis represents
projected separation $R$, rather than 3D separation $\rtd$.

The solid lines $\xint/b^2$ are identical to the matter correlation
function $\xi_m$ and the linear bias factor $b$ is constant. However,
the nonlinear evolution and galaxy formation process complicate
the relation between galaxies and underlying matter fluctuations, and 
galaxy bias becomes scale-dependent on small scales, even when the nonlinear
matter power spectrum is used (e.g., \citep{ZEZE05,YTWZKD06}).
While we plot the correlation functions at $\rtd\simeq0.5-200\hmpc$ for
completeness, the validity of our calculation is limited to the linear regime.
The solid lines at $\rtd=151~\mpc$ ($=106\hmpc$)
show prominent enhancement in the clustering amplitude, 
known as the baryonic acoustic peak \citep{PEYU70,SUZE70b}.
The baryon-photon plasma in the early universe propagates as sound waves
and these periodic oscillations in Fourier space translate into one peak in 
real space with its width deviating from a sharp delta function due to the 
termination of the harmonic series,
determined by the horizon size at the cosmological recombination epoch.
Note that the correlation function becomes negative at $\rtd\simeq128\hmpc$,
beyond which we plot its absolute value.

The correlation functions ($\xizz$, $\xidz$, and 
$\xizd$) of the redshift-space distortion bias have the overall
shape similar to $\xint$. However, since 
$\xizz$, $\xidz$, and $\xizd$
in Eqs.~(\ref{eq:pzz}) and~(\ref{eq:xidz})
have additional functional dependence on spherical Bessel functions $j_2(x)$ 
and $j_4(x)$ compared to $\xint$ in Eq.~(\ref{eq:xint}),
it puts more weight on higher $k$ and hence
the nonlinear effects persist up to 
$\rtd\simeq10\hmpc$ in Fig.~\ref{fig:corr}$a$,
larger than $3\hmpc$ for $\xint$. However,
the incoherent superposition of the additional Bessel
functions washes out the acoustic peak in the correlation functions of the
redshift-space distortion bias,
leaving little structure in $\xizz$, $\xidz$, and $\xizd$ at the acoustic
scale. Since the observed correlation function $\xiobs$ is
the sum of all the contributions and it is hard in practice to separate
each contribution from one another, it may 
look as if $\xint$ is swamped by $\xizz$, $\xidz$ and $\xizd$
at the acoustic scale, but note that we plot $\xint/b^2$ and
$\xidz/b$: the linear bias factor 
of luminous red galaxies is $b_0\simeq1.5-2.0$ 
\citep{EIBLET05,TEEIST06,PASCET07,BLCOET07}.
Assuming that galaxies
have no velocity bias $\bdv{v}_g=\bdv{v}$, the linear bias factor
at high redshift is $b(z)-1=(b_0-1)/D(z)$, sufficient for $\xint$
to show its structure, when combined with $\xizz$, $\xidz$, and $\xizd$, yet
the plot without $b$ captures the main structure of the correlation 
functions, since the linear bias factor is still of order unity.

Note that since the source galaxies are at the same redshift
in Fig.~\ref{fig:corr}$a$ to Fig.~\ref{fig:corr}$c$, the cosine angle of the
comoving separation is $\gamma=(\chi_2-\chi_1)/\rtd=0$, i.e., the
redshift-space correlation function (the sum of the solid, dotted, and dashed 
lines) is $\xi_\up{z-dist}=\xint+\xizz+\xizd+\xidz=
\xi_0-(1/2)~\xi_2+(3/8)~\xi_4$, 
different from the angle-averaged
(monopole) correlation $\xi_0$ often used in the literature 
\citep{EIZEET05,ROSHET08}. 
Figure~\ref{fig:comp} illustrates the multipole components of the 
redshift-space correlation function at $z=0.35$. The monopole ($solid$) is
identical to $\xint$ in shape but differs in normalization by 
$\xi_0=\xint\cdot c_0$ with the multipole coefficient $c_0$ in
Eq.~(\ref{eq:mulc}). The quadrupole $\xi_2$ ($dotted$) is negative by the
sign convention and the hexadecapole $\xi_4$ ($dashed$) is positive in the
figure, while the monopole $\xi_0$ changes its sign as $\xint$ changes at
$\rtd\gtrsim128\hmpc$ (see the inset). As noted before, the spherical Bessel
functions $j_2(k\rtd)$ and $j_4(k\rtd)$ in $\xi_2$ and $\xi_4$ peak at scales
different from the typical scale $k\sim1/\rtd$ for $\xi_0$, and thus
the acoustic structure seen in $\xi_0$ is smoothed out in $\xi_2$ and $\xi_4$.

In practice, galaxy
redshift surveys have a narrow but nonzero radial window function and galaxy
pairs in the same redshift bin often have the line-of-sight separation 
comparable to the transverse separation, i.e., $\gamma\ne0$. 
The angular dependence of the redshift-space correlation function, 
therefore, complicates the interpretation of its
measurements, which are further plagued by low
signal-to-noise ratios in estimates of $\xi_2$ and $\xi_4$. While the 
monopole $\xi_0$ can be used to ease the 
theoretical and/or observational challenge, full analysis of the anisotropic
structure of
$\xi_\up{z-dist}$ could in principle bring more information than  $\xi_0$ 
measurements (see \citep{OKMAET08,GACAHU08} for recent analysis). 
We analyze the full anisotropic structure of the observed correlation function
$\xiobs$ below. On small scales,
virial motions of galaxies result in additional anisotropic
structure in $\xi_\up{z-dist}$, known as the Finger-of-God (FoG) effect. 
Note that since this effect involves galaxy motions in nonlinear objects,
it is not considered in our calculation and
linear theory provides an inaccurate description of the FoG effect:
while a simple dispersion model \citep{BAPEHE96} is often adopted to extract 
additional information contained in the anisotropic structure,
it is demonstrated \citep{SCOCC04} that this model leads to an unphysical 
distribution of pairwise velocities. However, this difficulty could 
be tackled by recent approach based on modeling nonlinear galaxy bias 
in redshift-space
\citep{SELJA01,WHITE01,TINKE07}.

The short dashed and short dot-dashed lines in Fig.~\ref{fig:corr}
show the correlation functions
of the magnification bias $\xiMB/(5p-2)^2$ and the evolution bias
$\xievo/E^2$, respectively. The magnification bias $\xiMB$ is typically
smaller than $\xint$ by the ratio of the transverse correlation scale
$1/k_\perp$ to the Hubble distance $1/H$, and the magnification bias factor 
is of order unity, $(5p-2)=-1.0\sim2.0$ \citep{SCMEET05} for galaxies and
quasars, while it can be further suppressed by the source effect canceling the 
volume effect $(5p-2)\simeq0$. Since $\xiMB$ in Eq.~(\ref{eq:xiMB})
is proportional to the projected correlation function $w_p$, its overall
shape is similar to $\xint$ but $\xiMB$ is positive due to
projection of $\xint$. 
As the source population is located at higher redshift, 
longer line-of-sight distance increases the gravitational lensing effect
and $\xiMB$ increases in redshift,
as opposed to $\xint\propto D^2\propto 1/(1+z)^2$ 
decreasing in redshift. For example,
$\xiMB$ at $z=2.5$ in Fig.~\ref{fig:corr}$c$ can grow up to a few percent
of $\xint$ at the acoustic scale \citep{HUGALO07,VADOET07}.

\begin{figure}
\centerline{\psfig{file=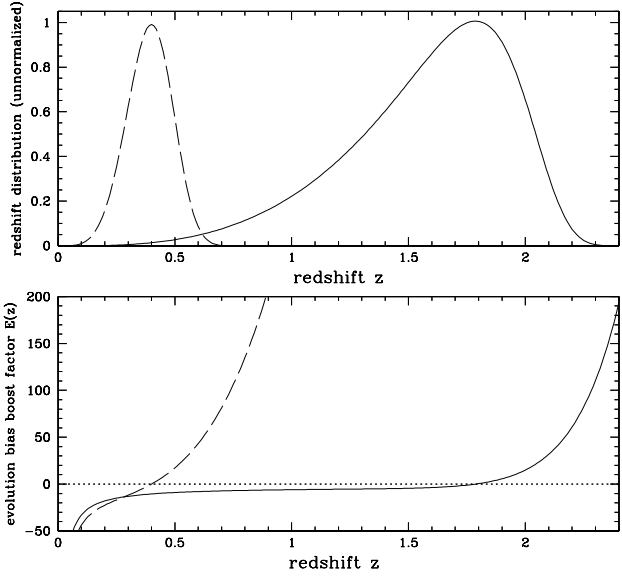, width=3.5in}}
\caption{Redshift distribution of source populations and boost factor
$E(z)$ of the evolution bias. Assuming the standard functional form
$\bar n(z)dz\propto z^\alpha\exp\left[-(z/z_0)^\beta\right]dz$, the redshift
distribution of galaxies ($dashed$) and quasars ($solid$) are shown in the
upper panel with $(\alpha,\beta,z_0)=(4,4,0.4)$ for galaxies and $(3,13,2)$
for quasars, 
respectively \citep{SCMEET05,JIFAET06,PASCET07}. The bottom panel shows the
evolution bias boost factor computed by using Eq.~(\ref{eq:eboo}).
The number density of sources changes exponentially beyond the mean
redshift and the evolution bias is substantially enhanced 
in proportion to $E(z)$.}
\label{fig:evo}
\end{figure}

The evolution bias $\devo$ is often ignored in the literature compared to the
redshift-space distortion bias $\zdist$, since 
$\devo\propto\varepsilon\simeq V$ and $V\ll\zdist$. 
However, the evolution bias can be
significantly enhanced when the mean number density of sources changes 
rapidly in redshift. To estimate the
evolution boost factor $E(z)$ in Eq.~(\ref{eq:eboo}), we assume
the standard functional form of a redshift distribution
$\bar n(z)dz\propto z^\alpha\exp\left[-(z/z_0)^\beta\right]dz$ and take
two source populations as illustrative examples: galaxies and quasars
characterized by $(\alpha,\beta,z_0)=(4,4,0.4)$ and $(3,13,2)$, respectively
\citep{SCMEET05,JIFAET06,PASCET07}. While the bright samples of luminous red
galaxies in the BOSS will have a redshift distribution flatter than the 
assumed here, the faint samples with larger number density and volume 
will have a non-flat redshift distribution \citep{EIANET01,ZEEIET05}. 
The clustering of Lyman-$\alpha$ forests at $z=2.5$ will 
be measured by the spectrum of quasars 
at $z>2.5$, not by quasars themselves at $z=2.5$. 
However, we simply assume that $\xiobs$ is measured from the galaxy samples
at $z=0.35$ and $z=0.6$, and from the quasar samples at $z=2.5$.

The upper panel of Fig.~\ref{fig:evo} illustrates the redshift distribution
of the galaxy ($dashed$) and quasar ($solid$) samples, with its peak at 
$z=0.4$ and~1.8, and the bottom panel shows the evolution boost factor $E(z)$
of each sample. For the assumed redshift distribution, the evolution boost
factor is typically a factor $\sim10$, and it vanishes at the peak redshift. 
However, a sharp decline in the mean number density of source populations
beyond the mean redshift makes the evolution bias $\devo$ sensitive to the
change in observed redshift $\zobs$ due the generalized Sachs-Wolfe effect,
and $E(z)$ can be further enhanced by another factor of ten.
With significant boost of
$E^2(z)\simeq100-10000$, the correlation $\xievo$ 
of the evolution bias
should be given a proper consideration, especially when the mean
number of the source population changes rapidly.
Note that the evolution bias
$\xievo/E^2$ ({\it short dot-dashed}) in Fig.~\ref{fig:corr}$a$ is
comparable to the magnification bias $\xiMB/(5p-2)^2$ ({\it short dashed})
and is larger at the acoustic peak scale, and the evolution boost factor $E(z)$
can be significantly larger than the magnification bias factor $(5p-2)$.
Therefore, it is of particular importance 
to select samples of source populations that
have relatively flat redshift distribution in number density ($E\simeq10$),
and to limit the redshift range of measurements below the peak redshift.
The short and the long dot-dashed lines in Fig.~\ref{fig:comp} show the 
multipole components $\xievo^l$ of the evolution bias, defined as
$\xievo=\sum_{l=0,2}\xievo^l(\rtd)P_l(\gamma)$. Both components are
smooth and change little over $\rtd=1-200\hmpc$.
With $\xievo\propto H^2a^2D^2\propto1/(1+z)$, 
it decreases slowly in redshift,
and the nonlinear effect is relatively small compared to $\xint$, 
since less weight is given to short wavelength modes. 

\begin{figure*}
\centerline{\psfig{file=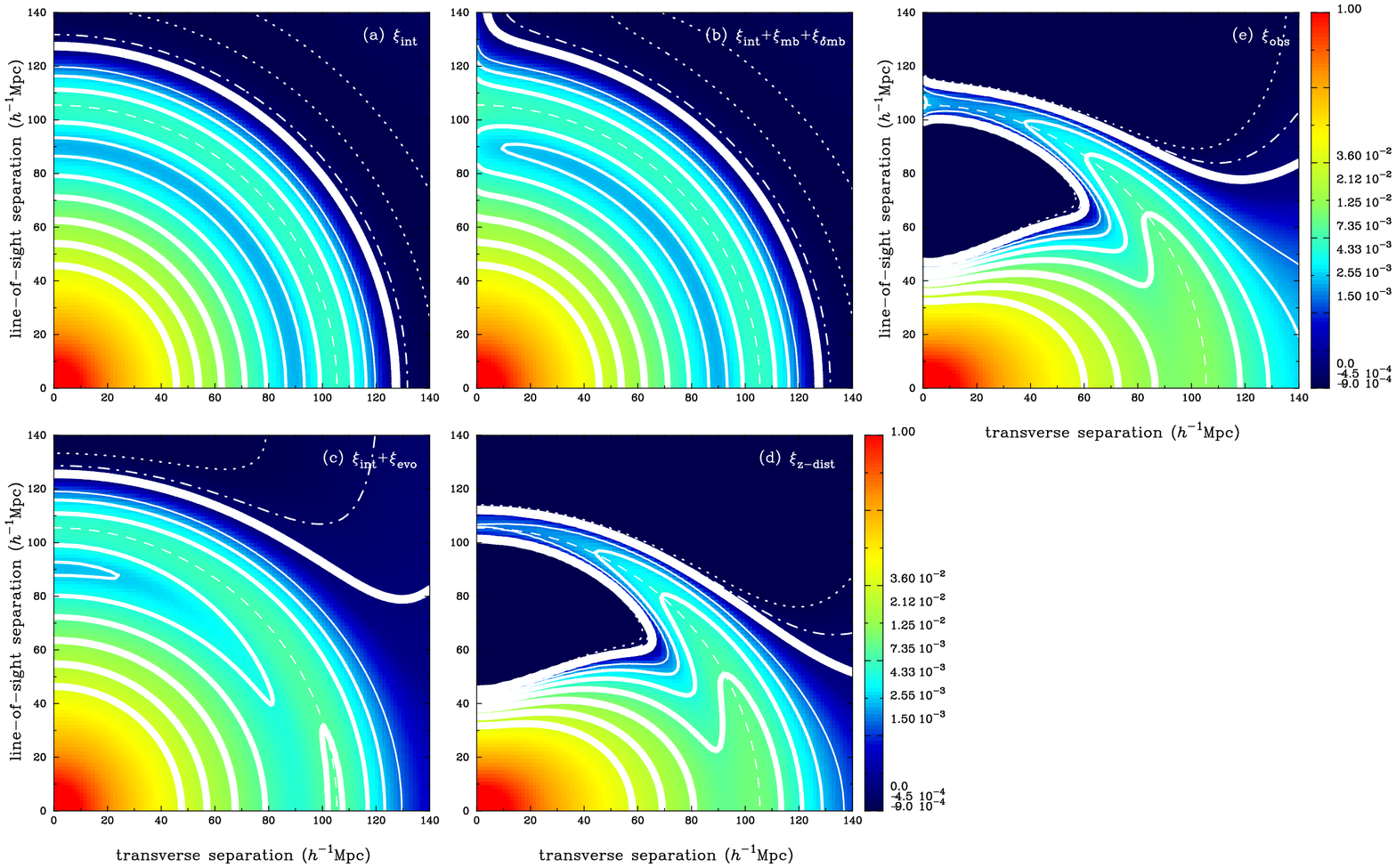, width=7.0in}}
\caption{(color online)
Anisotropic structure of the observed correlation function $\xiobs$
at $z=0.35$. We plot the correlation function $\xint$
of the intrinsic source fluctuations
in Panel~($a$), and show the individual effects of the magnification bias
$\xint+\xiMB+\xidMB$ ({\it Panel~b}), the evolution bias $\xint+\xievo$
({\it Panel~c}), and the redshift-space distortion bias
$\xi_\up{z-dist}=\xint+\xizz+\xizd+\xidz$
({\it Panel~d}) on the anisotropic structure.
The observed correlation function $\xiobs$,  
the sum of all the correlation functions, is shown in Panel~($e$). 
We assume that the galaxy bias factor
is $b=2$, the magnification bias factor is $(5p-2)=2$, and the evolution 
boost factor is $E=100$. The color maps are linearly proportional to the value
of correlation function $\xi$ in each panel
at $\xi<4\times10^{-4}$ and are logarithmically proportional
to $\xi$ at $\xi>4\times10^{-4}$.
The solid contours are logarithmically spaced at
$\xi\geq1.5\times10^{-3}$ and their thickness increases with the value of $\xi$
(contour values are labeled in the color bar), while the thickest solid
curves represent the $\xi=0$ contour. The dot-dashed and dotted curves 
represent the contours with $\xi=-4.5\times10^{-4}$ and $-9.0\times10^{-4}$, 
respectively. The acoustic scale in $\xint$ ($\rtd=106\hmpc$) 
is shown as dashed lines in each panel for reference.
Note that two regions are underrepresented in the color maps, as
the region with $\xi>1.0$ is highly concentrated at $\rtd\ll20\hmpc$
and there is no distinctive feature in the anisotropic structure around
the region with $\xi<0$.}
\label{fig:aniso}
\end{figure*}

Now recall that there
are four terms of $\zdist$ in Eq.~(\ref{eq:zfive}) that are ignored in our
calculation, and they are comparable to $\varepsilon\simeq V$, albeit smaller
than the dominant term in 
$\zdist\simeq-{1+z\over H}{\partial V\over\partial\chi}$. 
While there is no additional boost factor like $E(z)$ in $\devo$, 
their contributions
to $\xiobs$ are typically of order $\xievo/E^2$ and are as large as
$\xiMB$ at $z=0.35$ in Fig.~\ref{fig:corr}$a$.
Though $\xiMB$ is larger at higher redshift, their impact on $\xiobs$ 
also increases in redshift: approximately at the sub-percent level for each 
contribution at $z=2.5$. This level of accuracy would be appropriate given the 
statistical errors present in current samples, but further calculations of
the ignored terms may be needed in future surveys.

In Fig.~\ref{fig:corr}$d$, we consider the correlation functions of two
source populations, separately located
at $z_1=0.35$ and $z_2=0.6$ as a function of 2D projected 
separation $R=r(\bar\chi)\Delta\theta$.
Note that given a large line-of-sight separation $\sim600\hmpc$
between $z_1$ and $z_2$, all the correlation functions that depend on 3D
comoving separation $\rtd$ are nearly zero, i.e.,
$\xint\simeq\xi_{\delta\up{z}}\simeq\xi_{\up{z}\delta}\simeq\xizz\simeq
\xievo\simeq0$. The two non-vanishing contributions in 
Fig.~\ref{fig:corr}$d$ are the auto-correlation of the magnification bias
$\xiMB/(5p_1-2)(5p_2-2)$ ({\it short dashed}) and the cross-correlation of the
intrinsic source fluctuation and the magnification bias 
$\xidMB/b_1(5p_2-2)$ ({\it long dot-dashed}) that depend on the
projected separation, rather than 3D separation itself. Note that the 
cross-correlation in Fig.~\ref{fig:corr}$a$ to Fig.~\ref{fig:corr}$c$ is
identically zero: $\xidMB=0$ at $z_1=z_2$ with the Limber 
approximation we adopted here, but it is in general smaller than $\xiMB$
unless $z_1\neq z_2$ (see \citep{VADOET07} for a somewhat different derivation).
Note that while both $\xiMB$ and $\xidMB$ in 
Eqs.~(\ref{eq:xiMB}) and~(\ref{eq:xidmb}) depend on the projected separation
via $w_p$, $\xidMB$ has additional linear dependence on the comoving
line-of-sight separation $\chi_2-\chi_1=\Delta\chi$, and it increases with
$\Delta\chi$, as opposed to $\xiMB$ with little dependence on $\Delta\chi$.
With the large $\Delta\chi\sim600\hmpc$ in Fig.~\ref{fig:corr}$d$,
$\xidMB$ is substantially larger than $\xiMB$.

Figure~\ref{fig:aniso} examines the anisotropic structure of the observed
correlation function $\xiobs$, evaluated at $\bar z=0.35$. 
The $x$-axis represents the transverse separation $R=r(\bar\chi)\Delta\theta$
and the $y$-axis represents the line-of-sight separation 
$\Delta\chi=\chi_2-\chi_1$ with fixed $\bar z=(z_1+z_2)/2=0.35$, 
$\bar\chi=(\chi_1+\chi_2)/2=980\hmpc$, and $\chi(z_1)\leq\chi(z_2)$.
The color maps are linearly proportional to the value of the correlation 
function $\xi$ plotted in each panel below the adopted threshold
$\xi=4\times10^{-4}$, and they are logarithmically proportional
to $\xi$ above the threshold. The solid contours are also logarithmically 
spaced with increasing thickness at $\xi\geq1.5\times10^{-3}$ to emphasize
the structure shown as the color maps, and their contour values are labeled in 
the color bar. The thickest solid contours separate the regions with $\xi>0$
from those with $\xi<0$, and the dot-dashed and dotted curves represent the
contours with
$\xi=-4.5\times10^{-4}$ and $-9.0\times10^{-4}$, respectively. For reference,
we also plot the acoustic scale in $\xint$ ($\rtd=106\hmpc$) 
as dashed lines in each panel.
In Fig.~\ref{fig:aniso}$a$, we plot the correlation function $\xint$ of the
intrinsic source fluctuations, assuming the galaxy bias factor $b=2$.
As the rings of the concentric contours show, $\xint$ is spherically 
symmetric and depends only on 3D separation $\rtd$. 
The acoustic peak shows its structure as a circular ring
at $\rtd=106\hmpc$ ({\it dashed}) and beyond $\rtd\sim128\hmpc$ $\xint$ 
becomes negative without further distinctive feature in its structure.

The gravitational lensing effects, the magnification bias $\xiMB$ and 
its cross-correlation $\xidMB$, break the spherical symmetry in 
$\xint$, and its impact on the anisotropic structure is shown in 
Fig.~\ref{fig:aniso}$b$, assuming the magnification bias factor $(5p-2)=2$. 
$\xiMB$ depends on the line-of-sight separation $\Delta\chi$ only
through $\chi_1$ and $\chi_2$ in Eq.~(\ref{eq:xiMB}), and
$\Delta\chi$ is small compared to the line-of-sight distance, i.e.,
$\Delta\chi\ll\chi_1\simeq\chi_2$. Thus $\xiMB$ is virtually
independent of $\Delta\chi$ and is just a function of transverse separation
$R$, decreasing with increasing $R$. 
As is seen in Fig.~\ref{fig:corr}$a$, $\xiMB$ is
in general orders of magnitude smaller than $\xint$ at $z=0.35$,
but $\xint$ becomes negative at large $\rtd$ and smaller than $\xiMB$, e.g., 
$\xint=-8.4\times10^{-4}<\xiMB=4.8\times10^{-6}$ at $R=10.0\hmpc$ and
$\Delta\chi=140\hmpc$ ($\rtd=140.3\hmpc$). Since $\xiMB$ changes slowly with
$R$, the demarcation curve between the regions with $\xiMB>\xint$ and
$\xiMB<\xint$ roughly corresponds to the $\xint=0$ contour 
({\it thickest solid}) in
Fig.~\ref{fig:aniso}$a$. However, since $|\xint|>\xiMB$ in general
except at a narrow strip around the $\xint=0$ contour, the impact of
$\xiMB$ on the anisotropic structure is negligible at $\bar z=0.35$.
Note that the impact of $\xiMB$ is substantially enhanced at higher redshift,
where longer line-of-sight distance results in more fluctuations 
and the clustering amplitude of $\xint$ is lower.

While both $\xiMB$ and $\xidMB$ are proportional to $w_p$, $\xidMB$ depends
on $w_p$ itself, rather than the integral of $w_p$ along the line-of-sight,
on which $\xiMB$ depends:
$\xidMB$  becomes negative at large transverse separation 
$R\simeq110\hmpc$ as $\xint$ becomes negative at large 3D separation $\rtd$. 
Note that for an observable angular separation $\Delta\theta$, the transverse
separation $R=r(\bar\chi)\Delta\theta$ in the figure is slightly different
from $r(\chi_1)\Delta\theta$ for $w_p$ in Eq.~(\ref{eq:xidmb}), and this
difference tilts otherwise a vertical line with $\xidMB=0$ at $R\simeq110\hmpc$
toward larger $R$ at large $\Delta\chi$ ($\xidMB=0$ at $R\simeq128\hmpc$
and $\Delta\chi=140\hmpc$), because $\Delta\chi=\chi_2-\chi_1$ and
$z_1\le z_2$ with fixed $\bar z=(z_1+z_2)/2=0.35$.
Furthermore, since $\xidMB$ linearly
increases with $\Delta\chi$ (hence it vanishes at $\Delta\chi=0$), the
absolute value of $\xidMB$ is larger than $\xiMB$ except at the regions
at $\Delta\chi=0$ and around the nearly vertical strip with $\xidMB=0$, and
it is also comparable to $\xint$ at large $\Delta\chi$ and small $R$, e.g., 
$\xidMB=4.3\times10^{-4}$ at $R=10\hmpc$ and $\Delta\chi=140\hmpc$.
Therefore, when $\xint$ is combined with $\xiMB$ and $\xidMB$ as
shown in Fig.~\ref{fig:aniso}$b$,
$\xiMB$ has little impact but $\xidMB$ distorts the symmetric contours of
$\xint$ at large $\Delta\chi$ and small $R$. At higher redshift, the amplitude
of $\xidMB$ decreases with that of $\xint$, and $\xiMB$ becomes a more
dominant contribution than $\xidMB$.
Note that the amplitude at the acoustic scale ($dashed$) is not significantly
altered by
the gravitational lensing effects, even along the line-of-sight 
direction at $\bar z=0.35$.

The impact of the evolution bias $\xievo$ is shown in Fig.~\ref{fig:aniso}$c$,
where we assume the evolution boost factor $E=100$.
While $\xi^0_\up{evo}$ and $\xi^2_\up{evo}$
change slowly with separation, they have the opposite sign as shown in 
Fig.~\ref{fig:comp}. Therefore, they tend to cancel out along
the line-of-sight direction ($\gamma=1$), reducing the amplitude of $\xievo$,
and the largest contribution of $\xievo$ arises along the
transverse direction ($\gamma=0$), where the absolute values of the
monopole $\xievo^0$ and the quadrupole $\xievo^2$ add up. For example, 
$\xievo=1.6\times10^{-3}>\xint=-8.4\times10^{-4}$ at $R=140\hmpc$ and 
$\Delta\chi=10\hmpc$. As noted in Fig.~\ref{fig:corr}$a$, the amplitude of
$\xievo$ with $E=100$ is smaller than $\xint$ at $\rtd<50\hmpc$ and 
its impact is appreciable only at $\rtd\geq100\hmpc$ 
along the transverse direction.
As the angular separation becomes small with fixed 3D separation, the
impact of $\xievo$ decreases, because the second
order Legendre polynomial is a monotonic function of angle. Note that 
compared to $\xint$, $\xievo$ changes slowly in redshift and its impact is
larger at higher redshift for a fixed $E$. The overall shape of the acoustic
scale ($dashed$) also remains unaffected by the evolution bias.

\begin{figure}
\centerline{\psfig{file=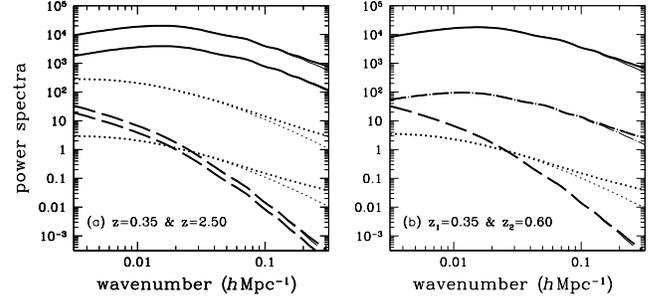, width=3.5in}}
\caption{Dissection of the observed galaxy power spectrum $P_\up{obs}$.
Solid and dashed 
lines represent power spectra of the intrinsic source fluctuations 
$P_\up{int}/b^2$ and the evolution bias $P_\up{evo}/E^2$. Power spectrum
of the magnification bias $P_\up{mb}/(5p-2)^2$ is shown as dotted
lines for a survey window of width $200\hmpc$ (see the text). 
Note that we omit power spectra of the 
redshift-space distortion bias $P_\up{zz}$, $P_{\delta\up{z}}$, and
$P_{\up{z}\delta}$, since they have the same shape as $P_\up{int}$
for the line-of-sight component of the
wavenumber $k_z=k$ up to numerical factors of order unity. Thin and thick 
lines represent power spectra computed by using the linear and nonlinear
matter power spectrum. {\it Left panel:} power spectra are computed at two
different redshifts, only the power spectrum of the magnification bias
increases at higher $z$, while $P_\up{int}$ and $P_\up{evo}$ decrease.
{\it Right panel:} cross-power spectra are computed
for two galaxy populations at $z_1=0.35$ and $z_2=0.6$, and dot-dashed
lines show the cross-power spectrum of the intrinsic source fluctuation
at $z_1$ and the magnification bias of source galaxies at $z_2$.
Large line-of-sight separation $\sim600\hmpc$ corresponds to 
$k_z\simeq0.002\hmpci$ and it has little impact on power spectrum 
($k\simeq k_\perp\gg k_z$).}
\label{fig:pow}
\end{figure}

Figure~\ref{fig:aniso}$d$ examines the redshift-space correlation function
$\xi_\up{z-dist}=\xint+\xizz+\xidz+\xizd$. Though its angular structure is 
similar to $\xievo$, it differs in two aspects:
$\xi_\up{z-dist}$ has the additional hexadecapole $\xi_4$,
and the quadrupole $\xi_2$ becomes dominant over the monopole $\xi_0$ at 
$\rtd\simeq50\hmpc$ smaller than $\rtd\simeq200\hmpc$, where 
$\xievo^0\simeq\xievo^2$, shown in Fig.~\ref{fig:comp}. Therefore, the angular 
structure changes more dramatically than that seen in Fig.~\ref{fig:aniso}$c$.
Note that while the fourth
order Legendre polynomial is not a monotonic function
of angle, the hexadecapole $\xi_4$ is generally smaller than $\xi_0$ and
$\xi_2$, and its contribution is minor.
The contours exhibits the well known Kaiser effect \citep{KAISE87}
that coherent infall toward the overdense regions squashes the clustering
amplitude and the underdense regions inflate along the line-of-sight.
The large region with negative values at $\Delta\chi\simeq50-100\hmpc$ and
$R\leq60\hmpc$ is the characteristic feature of this effect, largely due to
the negative quadrupole $\xi_2$, and this structure has been recently 
measured with high signal-to-noise ratio \citep{OKMAET08,GACAHU08}. Even
along the line-of-sight direction, the monopole $\xi_0$ briefly takes over the
negative quadrupole around the acoustic scale, because the clustering 
amplitude is enhanced. Furthermore, while the clustering amplitude increases
with angle at the acoustic scale ($dashed$), its structure manifests itself
as ridges \citep{MATSU04}.

Figure~\ref{fig:aniso}$e$ puts together our discussion, 
plotting the observed correlation function $\xiobs$ at $\bar z=0.35$, 
the sum of all the correlation functions shown in each
panel. The redshift-space distortion affects the anisotropic structure by far
the most among the other effects considered here. The gravitational
lensing effects, mostly from $\xidMB$ at low redshift, become important
only at a small transverse but large line-of-sight separation. With the
same dependence on quadrupole, $\xievo$ follows the similar angular pattern
of $\xi_\up{z-dist}$, boosting their contributions along the transverse
direction, but its sole impact shows up at $R\geq110\hmpc$ due to the lower
amplitude. The clustering amplitude of $\xiobs$ at the acoustic scale 
($dashed$) is
no longer constant, nor a monotonic function of angle.
The peak location, imprinted in $\xint$ with local enhancement in clustering
amplitude, remains largely
unaffected as the gravitational lensing and generalized
Sachs-Wolfe effects distorts the anisotropic structure. However, note that
it is beyond our scope of the current investigation to what accuracy the
acoustic peak remains unaffected by these effects (see, e.g., 
\citep{EIWH04,SEEI05,EISEWH07,VADOET07,GUBESM07,HUGALO07,CRSC08,LOHUGA08}
for recent work on the robustness of the baryonic acoustic peak).

Figure~\ref{fig:pow} shows the equivalent dissection of the contributions
of the gravitational lensing and the generalized Sachs-Wolfe effects to the
observed galaxy power spectrum $P_\up{obs}$
in Fourier space. The solid, dashed, and dotted
lines represent $P_\up{int}/b^2$, $P_\up{evo}/E^2$, and $P_\up{mb}/(5p-2)^2$,
respectively. The power spectra are also computed by using the linear ($thin$) 
and the nonlinear ($thick$) matter power spectrum as in Fig.~\ref{fig:corr}.
In Fig.~\ref{fig:pow}$a$, the source galaxies are assumed to be 
at the same redshift ($z=z_1=z_2$), and the two sets of lines 
show the power spectra at 
$z=0.35$ and $z=2.5$, which decrease in redshift except that
$P_\up{mb}$ increases as we have seen $\xiMB$ increase in redshift.
The power spectrum of the intrinsic source fluctuations 
$P_\up{int}/b^2$ ($solid$)
exhibits two characteristic scales in its structure: a series of the acoustic
oscillations starting at  $k\simeq 0.085\hmpci$, and the turnover in the
overall shape at $k\simeq0.015\hmpci$ imprinted by
the horizon size at the matter-radiation equality $z=3300$.

For simplicity, the wavenumber is set equal to the line-of-sight
direction $k=k_z$ for plotting $P_\up{evo}$ and to the transverse direction 
$k=k_\perp$ for plotting $P_\up{mb}$. The power spectrum of the evolution bias
$P_\up{evo}/E^2$ ($dashed$) is typically many orders-of-magnitude smaller than
$P_\up{int}/b^2$ at $k\geq0.03\hmpci$ in Fourier space, but its contribution
can be at the few percent level of $P_\up{int}$ at the acoustic scale and 
comparable to $P_\up{int}$ at the matter-radiation equality scale,
with the evolution boost factor $E^2\simeq10000$.
Note that the power spectra of the redshift-distortion bias
$P_\up{zz}$, $P_{\delta\up{z}}$, and $P_{\up{z}\delta}$ are omitted in the figure,
because they have the same shape as
$P_\up{int}$ up to numerical factors of order unity when $k=k_z$.

To plot the power spectrum of the magnification bias $P_\up{mb}/(5p-2)^2$ 
($dotted$),
we replace $(2\pi)\delta^D(k_z)$ by a flat window function of width 
$200\hmpc$, typical value in redshift surveys, hence the dotted line de facto
delineates the angular power spectrum of the magnification bias
$(5p-2)^2~C^{\kappa\kappa}_{l=k_\perp r(\bar\chi)}$ with a dimensional
coefficient $r^2(\bar\chi)\times(200\hmpc)$  (see Appendix~\ref{app:limber}).
While the magnification bias is negligible at $z=0.35$, its effect increases
with larger line-of-sight distance at higher redshift, amounting to a few
percent at the acoustic scale and larger at the matter-radiation equality
scale at $z=2.5$. However, note that even with relatively
large contributions to $P_\up{obs}$, the shift in the peak positions can
be at the sub-percent level or smaller \citep{VADOET07,HUGALO07}.

Figure~\ref{fig:pow}$b$ plots the cross power spectra of two source 
populations at $z_1=0.35$ and $z_2=0.6$. As opposed to the correlation 
functions shown in Fig.~\ref{fig:corr}$d$, all the power spectra that depend
on 3D wavenumber remains virtually unaffected by the large line-of-sight
separation $\sim600\hmpc$, because it corresponds to very small wavenumber
$k_z\simeq0.002\hmpci$ and $k\simeq k_\perp\gg k_z$. The dot-dashed lines
show the cross power spectrum of the intrinsic source fluctuation and the
magnification bias $P_{\delta~\up{mb}}$. Since it is proportional to the 
line-of-sight separation, its contribution is larger than $P_\up{mb}$ in 
Fig.~\ref{fig:pow}$b$, but it is absent in Fig.~\ref{fig:pow}$a$.

The anisotropic structure of the observed power spectrum $P_\up{obs}$ has been
well studied with main focus on the effect of the redshift-space distortion
bias \citep{KAISE87}, and the redshift-space power spectrum $P_\up{z-dist}$
in Eq.~(\ref{eq:pzf}) has the multipole components that are identical in shape
but only differ in normalization. The evolution bias results in the similar
angular pattern: two multipole components that share its shape with
$P_\up{int}$ with different normalization. However, note that since the
magnification bias and its cross term are intrinsically 2D quantities,
their impact on the anisotropic structure of $P_\up{obs}$ is small, even
with realistic survey window functions \citep{HUGALO08}.

\section{Discussion}
\label{sec:discuss}
Galaxy two-point statistics, correlation function in real space and power
spectrum in Fourier space, have been extensively used in cosmology 
to characterize the underlying matter fluctuations.
We have presented a coherent theoretical framework based on the linearized
Friedmann-Lema{\^\i}tre-Robertson-Walker (FLRW) metric for computing the
gravitational lensing and the generalized Sachs-Wolfe effects. Within this
framework, the metric perturbations are sourced by the underlying matter
fluctuations, and they naturally give rise to perturbations in the observable
redshift of source galaxies and their angular position on the sky. The time
component of the photon geodesic equations can be used to show the former,
the generalized Sachs-Wolfe effect \citep{SAWO67} that generalizes the
standard redshift-space distortion by peculiar velocities in a cosmological
context, including the Sachs-Wolfe and the integrated Sachs-Wolfe effects.
The spatial components of the photon geodesic equations can be used to 
derive the latter, the gravitational lensing effect that includes the weak
lensing distortion, magnification, and time delay effects. This unified 
treatment provides a complete description of the relation between
these seemingly different effects and the underlying matter fluctuations.

Furthermore, it becomes transparent in this treatment how the gravitational
lensing and the generalized Sachs-Wolfe effects affect the observed 
fluctuation field of source galaxies. To the linear order in perturbations,
we have computed all the additional contributions to the intrinsic source
fluctuation, arising from the gravitational lensing and the generalized
Sachs-Wolfe effects. We can gain more insight on the impact of these effects
by separating them as two physically distinct origins: 
the volume and the source effects.
The former effect that involves the change of volume is independent of 
source galaxy populations and hence regardless thereof the volume effect
is always present in galaxy two-point statistics. By contraries,
the latter effect depends on the intrinsic properties of source galaxy
populations and may vanish for a certain population. All of
these contributions to the intrinsic source fluctuations result in numerous
additional auto and cross terms in the observed galaxy two-point statistics,
and therefore proper account should be taken into these additional terms 
in interpreting measurements of galaxy two-point statistics from
upcoming dark energy surveys. 

With the complete list of the contributions of the gravitational lensing and
the generalized Sachs-Wolfe effects, separated as two physically distinct
origins, we have identified several contributions in the volume effect
and one contribution 
in the source effect, which are ignored in the 
standard treatment: the evolution bias in the source effect arises from the
generalized Sachs-Wolfe effect, when the mean number density of sources
changes rapidly in redshift, and its impact on the observed galaxy two-point
statistics can be substantially larger than that of the gravitational lensing
magnification bias. 
The ignored contributions in the volume 
effect are typically of order peculiar velocities and hence they are 
subdominant, compared to the standard redshift-space distortion effect. 
However, their impact is comparable to the magnification bias at low redshift.
While the cross term of the magnification bias and the intrinsic source 
fluctuation is more important at low redshift than the contribution of
the magnification bias itself in the gravitational lensing effect, 
further calculations of the additional contributions associated with
the volume effect may be needed, if higher
accuracy of theoretical modeling is required from observation.

We have investigated the impact of the additional contributions to the
anisotropic structure of the observed galaxy two-point statistics, after
simplifying some of the contributions to the intrinsic source fluctuations.
The redshift-space distortion affects the observed galaxy two-point statistics
most, imprinting its well-known feature in the anisotropic structure 
\citep{KAISE87,STWI95,HAMIL98}.
The gravitational lensing effect is small but non-negligible at a percent
level, particularly
along the line-of-sight separation and at high redshift, since their 
contribution increases with longer line-of-sight distance to the source 
galaxies and the clustering amplitude of the intrinsic source fluctuations
decreases in redshift. The evolution bias has an angular pattern similar to
the redshift-space distortion, but its impact becomes appreciable,
only at fairly large transverse separation. While it is challenging to
analyze the observed anisotropic structure of galaxy two-point statistics,
its full analysis from upcoming dark energy surveys
can provide a great opportunity to separately
identify each contribution from the gravitational lensing and the generalized
Sachs-Wolfe effects, increasing the leverage to understand the underlying
physical mechanism.

However, we note that constraining the underlying cosmological model
will require not only accurate theoretical predictions, but also model 
fitting to measurements, which results in further distortion
in galaxy two-point statistics, known as Alcock-Paczy\'nski 
effect \citep{ALPA79}.
Furthermore, our current investigation has focused on the linear theory
predictions and its additional contributions: nonlinearity and scale-dependent
galaxy bias can affect our results, though its impact is expected to
be less than at the percent level
around the acoustic scale (see, e.g., \citep{EIWH04,SEEI05,EISEWH07}).
However, additional leverage can be gained by modeling scale-dependent
galaxy bias on nonlinear scales \citep{YOWEET08}.

\acknowledgments 
We acknowledge useful discussions with Jordi Miralda-Escud\'e.
We are very grateful to Matias Zaldarriaga and Liam Fitzpatrick
for discussions of the effect of volume distortion.
J.~Y. is supported by the Harvard College Observatory under the
Donald~H. Menzel fund.

\appendix

\section{2D and 3D Statistics}
\label{app:limber}
Here we derive the relation between 2D and 3D fluctuations and their
two-point statistics.
Consider a fluctuation field $\delta^\up{2D}(\Vang;z_s)$ on the sky from
a source population at $z_s$. In general, it can be expressed in terms of
the convolution of a window function $W(\chi)$ and its 3D fluctuation
$\delta^\up{3D}(\bdv{x})$
\beeq
\delta^\up{2D}(\Vang;z_s)=\int_0^\infty d\chi ~W(\chi_s-\chi)~\delta^\up{3D}
\left[r(\chi)\Vang,\chi;z\right].
\eneq
When the window function is appreciable only around $z_s$ representing a 
narrow selection function in redshift surveys, 
$\delta^\up{2D}(\Vang;z_s)\simeq\delta^\up{3D}(\bdv{x};z_s)$ with its functional
dependence
$\bdv{x}=\left[r(\chi_s)\Vang,\chi_s\right]$. However, contributions to 
$\delta^\up{2D}(\Vang;z_s)$ can come from the fluctuations 
$\delta^\up{3D}(\bdv{x};z)$ at $z<z_s$ and
$\delta^\up{2D}(\Vang;z_s)$ may be substantially different from 
$\delta^\up{3D}(\bdv{x};z_s)$, when the window function is broad. For 
example,
the convergence field $\kappa(\Vang;z_s)$ in Eq.~(\ref{eq:conv}) has
the window function
\beeq
W^\kappa(\chi_s-\chi)=\left({3H_0^2\over2}\OM\right)
{r(\chi_s-\chi)~r(\chi)\over a(\chi)~r(\chi_s)},
\label{eq:window}
\eneq
which peaks roughly at a half of $r(\chi_s)$.

In a sufficiently small patch of the sky, the Fourier mode of 
$\delta^\up{2D}(\Vang;z_s)$ is
\bear
\delta^\up{2D}_\bdv{l}(z_s)&=&\int d^2\Vang~e^{-i\bdv{l}\cdot\Vang}~
\delta^\up{2D}(\Vang;z_s) \\
&=&\int_0^\infty d\chi~ {W(\chi_s-\chi)\over r^2(\chi)} \nonumber \\
&\times&\int{dk_z\over2\pi}~e^{ik_z\chi}
~\delta^\up{3D}\left[k_z,\bdv{k}_\perp={\bdv{l}\over r};z\right], \nonumber
\enar
and its (angular) power spectrum is
\bear
\label{eq:2dps}
C_l(z_1,z_2)&=&\int{d^2\bdv{l}'\over(2\pi)^2}~\langle\delta^\up{2D}_\bdv{l}(z_1)~
\delta^{\up{2D}*}_{\bdv{l}'}(z_2)\rangle \\
&=&\int d\chi_a\int d\chi_b~{W(\chi_1-\chi_a)W(\chi_2-\chi_b)
\over r(\chi_a)^2} \nonumber \\
&\times&\int{dk_z\over2\pi}~e^{ik_z(\chi_a-\chi_b)}~
P\left[k_z,k_\perp={l\over r(\chi_a)};z_a,z_b\right] \nonumber \\
&=&\int d\chi ~{W(\chi_1-\chi)W(\chi_2-\chi)\over r^2(\chi)}
P\left[k={l\over r(\chi_a)};z\right]. \nonumber 
\enar
The last equality is obtained by adopting the Limber approximation, in which
fluctuations along the line-of-sight are smoothed out and only long wavelength
modes ($k_z\simeq0$) can contribute to the integral \citep{LIMBE54,KAISE92}.
With the Limber approximation,
the angular correlation function is
\bear
\label{eq:2dcc}
w(\Delta\theta;z_1,z_2)&=&\langle\delta^\up{2D}(\Vang_1;z_1)~\delta^\up{2D}
(\Vang_2;z_2)\rangle \\
&=&\int_0^\infty d\chi~W(\chi_1-\chi)W(\chi_2-\chi) \nonumber \\
&\times&w_p\left[r(\chi)\Delta\theta;z\right], \nonumber
\enar
where the projected correlation function is
\bear
w_p\left[R;z\right]&=&\int_{-\infty}^\infty dr_\parallel~
\xi\left[r=\sqrt{R^2+r_\parallel^2};z\right] \\
&=&\int_0^\infty{k~dk\over2\pi}P(k;z)J_0(kR).\nonumber
\enar

The angular correlation function and power spectrum in Eqs.~(\ref{eq:xiMB})
and~(\ref{eq:akk}) can be readily obtained by substituting the window
function $W^\kappa(\chi)$ for the convergence in Eq.~(\ref{eq:window})
with $W(\chi)$ in Eqs.~(\ref{eq:2dps}) and~(\ref{eq:2dcc}). The cross
correlation function and power spectrum in Eqs.~(\ref{eq:xidmb}) 
and~(\ref{eq:admb}) can be computed in a similar manner, since 
$W(\chi_s-\chi)=\delta^D(\chi_s-\chi)$ gives 
$\delta^\up{2D}(\Vang;z_s)=\delta^\up{3D}(\bdv{x};z_s)$.

In Sec.~\ref{sec:two}, we associated the angular power spectrum 
$C_l^{\kappa\kappa}$ to a 3D power spectrum to compare its impact with other
3D power spectra. A 3D fluctuation field can be constructed from 
$\delta^\up{2D}(\Vang;z_s)$ by
\bear
\delta(\bdv{k};z_s)&=&\int d^3\bdv{x}~e^{-i\bdv{k}\cdot\bdv{x}}~
\delta^\up{2D}(\Vang;z_s) \\
&=&\int d\chi_s ~r^2(\chi_s)~e^{-ik_z\chi_s}~\delta^\up{2D}_{\bdv{l}=\bdv{k}_\perp 
r_s}(z_s)\nonumber \\
&=&(2\pi)~\delta^D(k_z)~r^2(\chi_s)~\delta^\up{2D}_{\bdv{l}=\bdv{k}_\perp r_s}(z_s).
\nonumber
\enar
We assumed $\delta^\up{2D}_\bdv{l}$ is independent of $z_s$ in the last
equality. For high redshift source populations, this approximation is 
accurate, since the growth of the comoving angular diameter distance flattens
at high $z$ and it becomes nearly constant. Within this approximation
$r(\chi_1)=r(\chi_2)=r(\bar\chi)$,
the 3D power spectrum is anisotropic and it is related to the angular 
power spectrum by
\bear
P(k_z,k_\perp;z_1,z_2)&=&\int{d^3\bdv{k}'\over(2\pi)^3}~\langle\delta(\bdv{k};
z_1)~\delta^*(\bdv{k}';z_2)\rangle \\
&=&(2\pi)~\delta^D(k_z)~r^2(\bar\chi)~C_{l=k_\perp r(\bar\chi)}(\bar z). \nonumber
\enar
In practice, $\delta^D(k_z)$ need to be replaced by a survey window function 
\citep{HUGALO08}, but note that it is crucial to assume the independence of 
source redshift, when computing the power spectrum.
We also note that the Limber approximation breaks down when the radial
window function of a survey is narrow compared to the correlation length scale
(see, e.g., \citep{SIMON07,LOAF08}). However, the use of the Limber 
approximation is readily
justified in galaxy surveys, in which the radial window
function has width of $\Delta z\simeq0.1-0.2$, corresponding to several
hundred Mpc.

\section{Gauge-Invariant Form of Observed Number Density}
\label{app:ngal}
Here we provide a rigorous derivation of the observed number density
in Sec.~\ref{ssec:vol}. For simplicity, we assume a flat universe.

In the observer's frame, local coordinates $p^\alpha$ are used
to describe the observed positions of galaxies and their true positions are
related to the observed positions by 
the photon geodesic $x^a(\lambda)$.
The total number of observed galaxies can be computed by
considering a covariant volume integral \citep{WEINB72}:
\beeq
N_\up{gal}=\int\sqrt{-g}~n_\up{phy}~u^d~dS_d~,
\label{Aeq:ngal}
\eneq
where the (oriented) hyper-surface element is
\beeq
dS_d=\epsilon_{abcd}~
{\partial x^a\over\partial p^1}{\partial x^b\over\partial p^2}
{\partial x^c\over\partial p^3}~dp^1dp^2dp^3~,
\eneq
$g$ is the determinant of the space-time metric, $n_\up{phy}$ is the physical
number density of  galaxies, and $\epsilon_{abcd}=\epsilon_{[abcd]}$ 
is the Levi-Civita tensor density. We take the Newtonian gauge variables
$(\ztobs,\tobs,\pobs)$ as the observed local coordinates $p^\alpha$. 
For notational simplicity, tilde is used to represent observed quantities
(e.g., $\ztobs=\zobs$).
By imposing the number conservation, the observed number density is then
related to the total number of galaxies by
\beeq
N_\up{gal}=\int~\ntobs~{\chi^2(\ztobs)\over H(\ztobs)}~\sin\tobs
~d\ztobs d\tobs d\pobs~.
\label{Aeq:nobs}
\eneq

In a homogeneous universe, the local coordinates are identical to the
true coordinates $(\ztobs,\tobs,\pobs)=(z,\theta,\phi)$, and
the photon geodesic is simply
$x^a(\lambda)=(y,y~e^\alpha)$, where we choose the normalization of the affine
parameter as $\lambda=(a/\nu)y$. Noting that the four velocity of a comoving 
observer is $u^a=(1/a,0)$, equation~(\ref{Aeq:ngal})
can be readily solved as
\bear
N_\up{gal}&=&\int a^4~{n_\up{phy}\over a}~\epsilon_{\alpha\beta\gamma0}~
{\partial x^\alpha\over\partial \ztobs}{\partial x^\beta\over\partial \tobs}
{\partial x^\gamma\over\partial \pobs}~d\ztobs d\tobs d\pobs \nonumber \\
&=&\int a^3~n_\up{phy}~{\chi^2(z)\over H(z)}~\sin\theta~dz d\theta d\phi.
\enar
Therefore, we recover the standard relation for $N_\up{gal}$
and $\ntobs=a^3~n_\up{phy}=n(z,\theta,\phi)$. 

In an inhomogeneous universe, the photon geodesic deviates from the null path
and the local coordinates are different from the true coordinates.
Perturbations to the photon geodesic in an inhomogeneous universe
can be computed by integrating the null vector $k^a(\lambda)=dx^a/d\lambda$,
\beeq
x^a(\lambda)=(y,y~e^\alpha)+\int_0^ydy'~(\delta\nu,\delta e^\alpha)~.
\eneq
Note that to the first order in perturbations
the integration is performed
along the null path, ranging from the observer
at $y=0$ to the source galaxies at $y$.

With $u^a=((1-\psi)/a,v^\alpha/a)$, the integrand of equation~(\ref{Aeq:ngal})
is
\bear
u^d~dS_d&=&{1-\psi\over a}~\epsilon_{\alpha\beta\gamma0}~
{\partial x^\alpha\over\partial \ztobs}{\partial x^\beta\over\partial \tobs}
{\partial x^\gamma\over\partial \pobs}  \nonumber \\
&+&{v^\alpha\over a}~\epsilon_{abc\alpha}~
{\partial x^a\over\partial \ztobs}{\partial x^b\over\partial \tobs}
{\partial x^c\over\partial \pobs}~.
\label{Aeq:int}
\enar
The last two terms, proportional to $\psi$ and $v^\alpha$, contribute to the
first order in perturbations and the partial derivatives need to be computed,
only to the zeroth order. The first term has two sources of perturbations
from the partial derivatives: perturbations in the photon geodesic and
the relation between the local and true coordinates. The former is non-zero,
only when the derivative is taken with respect to $\ztobs$, i.e., 
\beeq
{1\over a}~\epsilon_{\alpha\beta\gamma0}\left(
{\partial x^\beta\over\partial \theta}{\partial x^\gamma\over\partial\phi}
\right)_0{\delta e^\alpha\over H(z)}~,
\eneq
and the latter is
\beeq
{1\over a}~\epsilon_{\alpha\beta\gamma0}\left({\partial x^\alpha\over\partial z}
{\partial x^\beta\over\partial \theta}{\partial x^\gamma\over\partial\phi}
\right)_0
\left({\partial z\over\partial\ztobs}+{\partial\theta\over\partial\tobs}
+{\partial\phi\over\partial\pobs}\right)_1~,
\eneq
where the subscripts denote the order in perturbations,
to which quantities in the bracket need to be computed.
When combined together, equation~(\ref{Aeq:int}) is
\bear
u^d~dS_d&=&{1\over a}{\chi^2(z)\over H(z)}\sin\theta\\
&\times&\left[1+\delta e^\alpha e_\alpha+\left(
{\partial z\over\partial\ztobs}+{\partial\theta\over\partial\tobs}
+{\partial\phi\over\partial\pobs}\right)_1
-\psi+v^\alpha e_\alpha\right]~. \nonumber
\enar

Finally, the determinant in equation~(\ref{Aeq:ngal}) gives
$\sqrt{-g}=a^4~(1+\psi+3~\phi)$ and the total number of observed galaxies is
\bear
N_\up{gal}&=&\int a^3~n_\up{phy}~
{\chi^2(z)\over H(z)}~\sin\theta ~d\ztobs d\tobs d\pobs \\
&\times&\left[1+2~\phi+\varepsilon+\left(
{\partial z\over\partial\ztobs}+{\partial\theta\over\partial\tobs}
+{\partial\phi\over\partial\pobs}\right)_1\right]~. \nonumber
\enar
From equation~(\ref{Aeq:nobs}), we obtain our final result,
\beeq
\ntobs=n~{\chi^2(z)\over\chi^2(\ztobs)}~{H(\ztobs)\over H(z)}~
\left[1+2~\phi-(1+z){d\varepsilon\over dz}-2~\kappa\right]~,
\eneq
which includes the distortion of volume element $\delta V$ and gravitational
lensing magnification. This expression is manifestly gauge-invariant and
valid on all scales. In special relativity, the volume element of the local
Lorentz frame of matter is distorted by $\gamma=\sqrt{1-v^2}$, and hence is
identical to an observer at rest, to the first order in perturbations.
However, in our case the volume element can be measured, only by observing
light rays of photons, of which time component is related to the spatial
component, giving rise to the first order distortion in volume element.

\bibliography{ms.bbl}

\end{document}